\documentclass[apj]{emulateapj}  
        \usepackage[ dvips, 
                 bookmarks,
                 bookmarksopen = true,
                 bookmarksnumbered = true,
                 breaklinks = true,
                 linktocpage,
                 hyperindex = false,
                 hyperfigures,
                 pagebackref,
                 colorlinks=true,
                 linkcolor=blue,
                 urlcolor=blue,
                 citecolor=red,
                 anchorcolor=green
                 ]{hyperref}{\usepackage[dvips]{hyperref}}
\usepackage{graphicx}
\usepackage{amsmath}
\usepackage{comment}
\usepackage{txfonts}
\usepackage{natbib}			

\usepackage[percent]{overpic} 

\renewcommand\email\texttt

\begin{document}
 
\slugcomment{\sc submitted to \it Astrophysical Journal}
\shorttitle{\sc The Distribution of the M31 Satellite System}
\shortauthors{Conn et al.}
 
\title{The Three-Dimensional Structure of the M31 Satellite System; \\
Strong Evidence for an Inhomogeneous Distribution of Satellites}

\author{A.\ R. Conn\altaffilmark{1, 2, 3}}

\author{G.\ F. Lewis\altaffilmark{4}}

\author{R.\ A. Ibata\altaffilmark{3}}

\author{Q.\ A. Parker\altaffilmark{1, 2, 5}}

\author{D.\ B. Zucker\altaffilmark{1, 2, 5}}

\author{A.\ W. McConnachie\altaffilmark{6}}

\author{N.\ F. Martin\altaffilmark{3}}

\author{D. Valls-Gabaud\altaffilmark{7}}

\author{N. Tanvir\altaffilmark{8}}

\author{M.\ J. Irwin\altaffilmark{9}}

\author{A.\ M.\ N. Ferguson\altaffilmark{10}}

\author{S.\ C. Chapman\altaffilmark{11}}


\altaffiltext{1}{Department of Physics \& Astronomy, Macquarie University, NSW 2109, Australia.}
 
\altaffiltext{2}{Research Centre in Astronomy, Astrophysics and Astrophotonics (MQAAAstro), Macquarie University, NSW 2109, Australia.}
 
\altaffiltext{3}{Observatoire astronomique de Strasbourg, Universit{\'e} de Strasbourg, CNRS, UMR 7550, 11 rue de l'Universit{\'e}, F-67000 Strasbourg, France}
 
\altaffiltext{4}{Sydney Institute for Astronomy, School of Physics, A28, University of Sydney, 
Sydney NSW 2006, Australia.} 
 
\altaffiltext{5}{Australian Astronomical Observatory, PO Box 296, Epping, NSW 2121, Australia.}
 
\altaffiltext{6}{NRC Herzberg Institute of Astrophysics, 5071 West Saanich Road, Victoria, British Columbia, Canada V9E 2E7.}

\altaffiltext{7}{Observatoire de Paris, LERMA, 61 Avenue de l'Observatoire FR 75014 Paris
France.}

\altaffiltext{8}{Department of Physics and Astronomy, University of Leicester, Leicester LE1 7RH, UK.}

\altaffiltext{9}{Institute of Astronomy, University of Cambridge, Madingley Road, Cambridge CB3 0HA, UK.}

\altaffiltext{10}{Institute for Astronomy, University of Edinburgh, Royal Observatory, Blackford Hill, Edinburgh EH9 3HJ, UK.}

\altaffiltext{11}{Department of Physics and Atmospheric Science, Dalhousie University, 6310 Coburg Road, Halifax, Nova Scotia, B3H 4R2, Canada.}

 
 
\begin{abstract}
	We undertake an investigation into the spatial structure of the M31 satellite system utilizing the distance distributions presented in a previous publication. These distances make use of the unique combination of depth and spatial coverage of the Pan-Andromeda Archaeological Survey (PAndAS) to provide a large, homogeneous sample consisting of $27$ of M31's satellites, as well as M31 itself. We find that the satellite distribution, when viewed as a whole, is no more planar than one would expect from a random distribution of equal size. A disk consisting of 15 of the satellites is however found to be highly significant, and strikingly thin, with a root-mean-square thickness of just $12.34^{+0.75}_{-0.43}$ kpc. This disk is oriented approximately edge on with respect to the Milky Way and almost perpendicular to the Milky Way disk. It is also roughly orthogonal to the disk like structure regularly reported for the Milky Way satellite system and in close alignment with M31's Giant Stellar Stream. A similar analysis of the asymmetry of the M31 satellite distribution finds that it is also significantly larger than one would expect from a random distribution. In particular, it is remarkable that $20$ of the $27$ satellites most likely lie on the Milky Way side of the galaxy, with the asymmetry being most pronounced within the satellite subset forming the aforementioned disk. This lopsidedness is all the more intriguing in light of the apparent orthogonality observed between the satellite disk structures of the Milky Way and M31.

\end{abstract}



 
\section{Introduction}
\label{s_intro}

The possibility that irregular distributions of satellite galaxies may be a common feature of large galaxy halos was originally bolstered by several studies of the anisotropic distribution of our own galaxy's satellites. \citet{LB1976} found that the Magellanic Stream along with Sculptor and the Draco-Ursa Minor Stream and their associated dwarf spheroidal galaxies all appear to lie in the orbital plane of the Magellanic Clouds. In \citet{LB1982}, all the then known dwarf spheroidal companions of the Milky Way are identified as lying in one of two streams. \citet{Kroupa05} examined the likelihood of producing the observed disk-like distribution of Milky Way satellites from a spherical or oblate dark matter halo. From comparisons with theoretical isotropic satellite distributions produced from such a halo, they find that the chance of producing the observed distribution from the dark-matter sub-halos of cold-dark-matter (CDM) cosmology is less than 0.5 \%. They examine various combinations of the inner most satellites and find a best-fit plane that is almost perpendicular to the plane of the Milky Way with a root-mean-square height ranging from only about $10$ to $30$ kpc. \citet{Zentner05}, whilst finding a similar plane to \citet{Kroupa05} for the Milky Way satellites, disagree with their assumption that such a plane is unlikely to arise from a conventional CDM dark matter halo. They argue that that the most luminous satellites cannot be taken for granted as forming randomly from the isotropic sub-halo distribution but instead, lie preferentially at smaller distances from the halo centre and co-planar with the major axis of the host halo. Coupled with the finding that galaxies preferentially align themselves with their major-axis highly-inclined or even perpendicular to that of the surrounding matter (e.g. \citealt{Navarro04}; \citealt{Hartwick2000}), this would provide a good explanation for the observed orientation of the best fit plane. The arguments of  \citet{Zentner05} are contended by \citet{Metz07} however, and it should be noted that in contrast to \citet{Zentner05}, \citet{Libeskind05} found that the distribution of the most massive subhaloes is not as flattened as the distribution of the Milky Way's satellites. 

	More recently, \citet{Lovell11}, using the six halo models in the Aquarius Simulations \citep{Springel08}, find that all six halos produce a significant population of sub-halos with quasi-planar orbits aligned with the main halo spin. This, they argue, is a natural explanation for the observed satellite distribution of the Milky Way. \citet{Paw12} argue against this however. With the calculation of the angular momenta of 8 Milky Way Satellites \citep{Metz08} revealing a strong alignment between 6 of the orbital poles, \citet{Paw12}  examine the likelihood of randomly drawing 6 sub-halos from each of the 6 Aquarius simulations (among other halo simulations), and finding a similar degree of alignment. More precisely, they draw $10^5$ sets of 8 satellites from each of the 6 simulations, and select the 6 with the highest degree of alignment between their orbits, thus emulating the findings of \citet{Metz08}. They then look at the degree of clumping of the orbital poles $\Delta_{sph}$ as well as the angular distance of the average of the orbital pole inclinations from the model equator $d$ and find that the actual degree of planarity observed for the six satellites identified by \citet{Metz08} ($\Delta^{MW}_{sph} = 35.4^{\circ}$ and $d_{MW} = 9.4^{\circ}$) are equalled or exceeded in the random draws in less than $10 \%$ of cases when $\Delta_{sph}$  is considered and less than $15 \%$ of cases for $d$. \citet{Starkenburg12} also find that the degree of planarity observed for the Milky Way satellites is uncommon in all six of the Aquarius halos (see Fig. 7 of that study).            

In addition to the revelation that the Milky Way's satellites appear to inhabit highly-inclined great planes, they also appear to corroborate the finding of \citet{Holmberg69}, namely that the companions of Spiral Galaxies preferentially congregate at  high galactic latitudes (the Holmberg Effect), as observed in his study of 174 galaxy groups. It is not clear why this should be the case, or even if it truly is the case, although if the apparent adherence of satellite systems to polar great planes is typical of galaxies in general, then the Holmberg Effect seems to be an extension of this. \citet{Quinn1986} proposed that dynamical friction may be responsible for the observed polar great planes, with those orbits spending the most time in close proximity to the galactic disk, experiencing the fastest decay, while those that take the most direct route through the disk environs, namely the polar orbits, experiencing the slowest orbital decay. It should be noted however that dynamical friction, whilst producing more polar orbits, would not produce planes of satellites. Nor would it be effective on the young globular clusters which are shown to be co-planar with the ``Vast Plane of Satellites'' identified around the Milky Way by \citet{Paw12B}. Indeed, \citet{Angus11} show that dynamical friction did not play a role in the formation of the Milky Way satellite orbits.

Besides the conjecture that satellite great planes trace the major-axis of the dark-matter halo in which the parent galaxy resides, there are other proposed mechanisms for their creation. One hypothesis is that these planes trace the orbits of ancient galaxies that have been cannibalized by the host galaxy. \citet{Palma02} have investigated this hypothesis by looking for planes among groups of satellite galaxies and globular clusters in the Milky Way's outer halo and find various members to be co-planar with either the Magellanic or Sagittarius streams. The findings of \citet{LB1995} are also consistent with such a hypothesis. Indeed, it is this hypothesis which is most strongly supported by \citet{Paw12}, wherein the $\Delta_{sph}$ and $d$ of satellites drawn from various tidal models equal or exceed $\Delta^{MW}_{sph}$ and $d_{MW}$ in over $80 \%$ of draws in some cases.  A similar hypothesis, which in some regards links the galaxy-cannibalization and dark-matter hypotheses, proposes that the observed planes result from the orientation of the large-scale filamentary structure of galaxy clusters (e.g. \citealt{Knebe04}), an orientation traced out by those minor galaxies which fall into the halo of a major galaxy. \citet{Metz09} argue however that extra-galactic associations of dwarf galaxies are too extended to account for the high degree of planarity observed for the Milky Way satellites. This argument is supported by the findings of \citet{Vera11} based on the Aquarius Simulations.

The great obstacle to a conclusive resolution of these issues is the lack of systems for which reliable spatial (and kinematic) data exists.   While some such data does exist for large galaxy clusters such as Virgo and Coma, accurate 3D distributions of galaxies within their halo have for a long time been known only for our own galaxy's halo, ascertainable due to our central position within it. It has only been in recent times that a second system has opened up to us - that of our counterpart in the Local Group, M31. Whilst various databases of photometry and other data have been available for M31 and some of its brighter companions for over a decade, it is the Pan-Andromeda Archaeological Survey (PAndAS - \citealt{McConn09}) - a deep photometric, 2-colour survey providing a uniform coverage of the M31 halo out to approximately 150 kpc - that has provided a new level of detail for this system. It is from this survey that we obtained our distances to M31 and 27 of its companions, following the method developed in \citet{Conn11} (henceforth CLI11) and further adapted for this purpose in \citet{Conn12} (henceforth CIL12).  The distances themselves and their associated uncertainty distributions are presented in CIL12 and it is these distributions that are utilized for all analysis contained in this paper.

With regard to previous studies of the anisotropy in the M31 satellite distribution, two investigations warrant consideration at this point. \citet{McConn06}, making use of Wide Field Camera (WFC) photometry from the Isaac Newton Telescope (INT) in what was essentially the forerunner to the PAndAS Survey, focus on ``Ghostly Streams" of satellite galaxies following a similar approach as \citet{LB1995} used for the Milky Way. In addition, they characterize the large degree of asymmetry in the satellite distribution, a feature also noted in CIL12, and examine the radial distribution of the satellites, noting a (statistically insignificant) larger average distance from M31 than that observed between the Milky Way and its satellites. They find a large number of candidate satellite streams, with some favoring the dwarf spheroidal members. \citet{Koch06} utilize distance measurements from a variety of sources and focus particularly on planes of satellites and, whilst they do not find a particularly significant best fit plane when their whole satellite sample is considered, it is rather interesting that they find a 99.7 \% statistical significance to their best fit plane when the then-known dwarf spheroidal galaxies dominate their sample. Furthermore, this plane is near-polar - as has been observed for the Milky Way, although they find little support for the Holmberg Effect. \citet{Koch06} utilize a particularly robust method in their search for high-significance planar fits to subsets of galaxies by considering every possible combination of a given number of satellites from their sample.

In the current study we employ a similar approach, but with the great advantage of having a considerably extended sample of galaxies in our sample, with all distances derived by the same method and from the same data as described in CLI11 and CIL12. (Distances are sampled from the distributions without the density prior applied - e.g. CIL12, Table 1, Column 2 - whilst the best fit distances are drawn from CIL12, Table 2, Column 4.) As a result, we are able to give full consideration to the effects of selection bias on the observed satellite distribution. This then presents an excellent opportunity to greatly improve our knowledge of the three-dimensional structure of the M31 satellite distribution, with important implications regarding the recent evolution of the system. 

A breakdown of the structure of the paper is as follows. In Section \S \ref{s_Method}, we outline our method for plane fitting (\S \ref{ss_Method_pf}) and locating significant planes of satellites as well as the orientation, magnitude and significance of the asymmetry of the distribution.  A method for generating random realizations of satellites subject to the same selection biases as the real data is also discussed in this section (\S \ref{ss_Method_rr}) as is the selection bias itself (\S \ref{ss_Method_sdb}). \S \ref{s_results} then presents the results of applying these methods, first to the sample as a whole, and then to subsets of galaxies. Specifically, \S \ref{ss_BFP_all_sats} presents a study of planarity within the satellite system when all satellites contribute to the determination of the best fit plane; \S \ref{ss_asy_all_sats} examines the asymmetry in a similar way;  \S \ref{sat_subsets} examines the orientations of planes of smaller subsets of satellites within the distribution; and \S \ref{Great_Plane} concludes this section with a determination of the significance of a `Great Plane' of satellites emerging from the preceding sections. Sections \ref{s_Discussion} and \ref{s_Conclusions} then follow with discussion and conclusions.  

Note that this paper was written in conjunction with a shorter contribution (\citealt{Ibata13}; hereafter ILC13) which announced some of the key discoveries resulting from the analysis we present here. In particular, the process of identifying the member satellites of the `great plane' discussed in ILC13 is described here in more detail. In this analysis however, we concern ourselves with the \emph{spatial} structure of the satellite system only and so the reader should refer to ILC13 for the interesting insight provided by the addition of the velocity information.


\section{Method}
\label{s_Method} 

\subsection{Plane Fitting}
\label{ss_Method_pf} 

In order to find planes of satellites within the M31 satellite system, our first concern is to convert the satellite distances as presented in CIL12 into three-dimensional positions. To do this, we begin with an M31-centered, cartesian coordinate system oriented such that the x and y axes lie in the M31 tangent plane with the z-axis pointed toward the Earth. Specifically, the x-axis corresponds to $\eta_{tp} = 0$ which is the projection of M31's Declination onto the tangent plane. The y-axis then corresponds to $\xi_{tp} = 0$ - the projection of M31's Right Ascension onto the tangent plane. The z-axis then points along the Earth-to-M31 vector, with magnitude increasing with distance from Earth. This orientation can be seen in Fig. 10(c) of CIL12. Thus:    

 \begin{equation}
 \begin{split}
    x& = D_{sat} \rm{cos}(\theta) \rm{tan}(\xi)\\ 
    y&=  D_{sat} \rm{sin}(\eta)\\
    z&=  D_{sat} \rm{cos}(\theta) - D_{M31} 
\end{split}
\label{e_cart_coord}
\end{equation}
where $D_{M31}$ and $D_{sat}$ are the distances from Earth to M31 and from Earth to the satellite respectively, $\theta$ is the angular separation on the sky between M31 and the satellite, and $\eta$ and $\xi$ are the real-angle equivalents of the tangent plane projection angles $\eta_{tp}$ and $\xi_{tp}$ respectively. 

\begin{figure*}[htbp]
\begin{center}
\includegraphics[width = 0.50\textwidth,angle=-90]{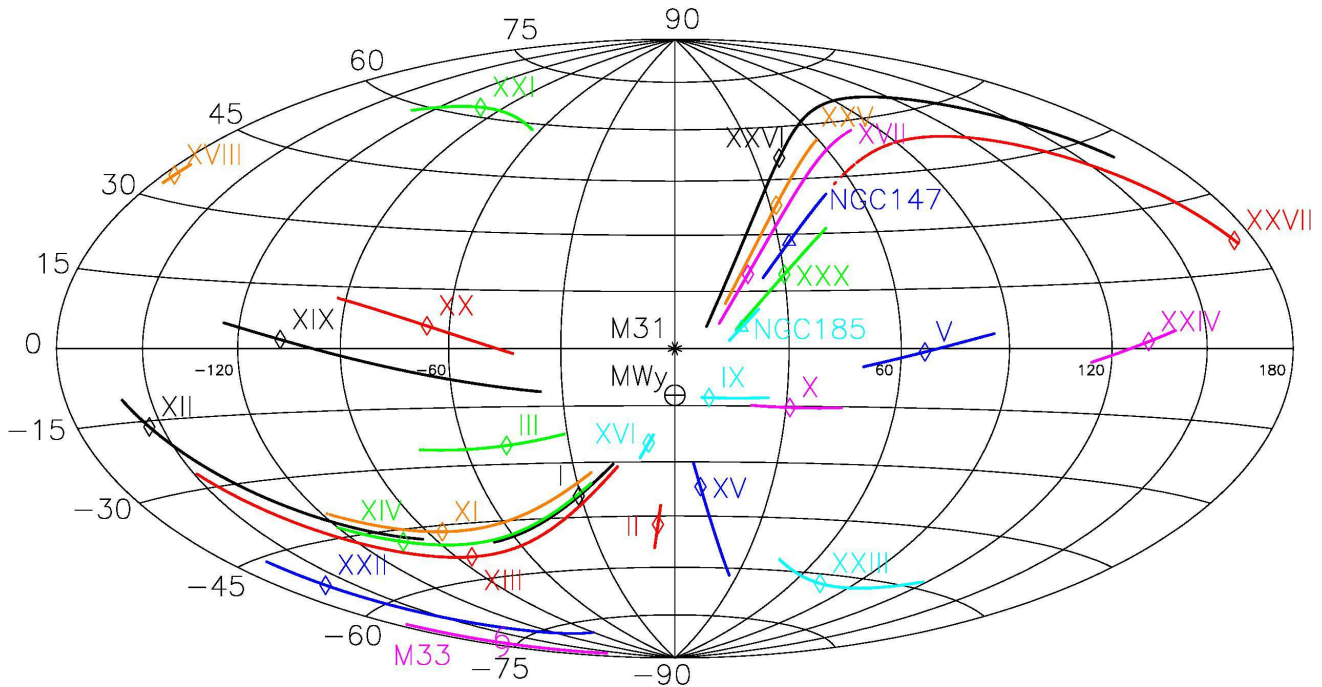}				  
\caption{An Aitoff-Hammer Projection showing the positions of M31's satellites, along with their associated $1 \sigma$ uncertainties. The positions show where each object would appear in the sky if viewed from the centre of M31, and are given in M31's inherent galactic latitude and longitude. The position of the Milky Way is also shown for reference. The position uncertainties trace single arcs across the M31 sky, rather than two-dimensional patches on account of the restriction of the uncertainty to lie solely along the line of sight to the object from Earth. For this same reason, all the lines point radially outward from the Milky Way. These uncertainties also take into account the uncertainty in M31's distance. The existence of a prominent plane, broadly consisting of Andromedas I, XI, XII, XIII, XIV, XVI, XVII, XXV, XXVI, XXVII, XXX and NGC147 and NGC185, is hinted at by the close proximity of their respective arcs.}
\label{sat_aitoff}
\end{center}

\end{figure*}

Next, we rotate this reference frame to the conventional M31 reference frame such that the positive z-axis points toward M31's north galactic pole\footnotemark[1] (i.e. $b_{M31} = +90^{\circ}$) and the $l_{M31} = 0^{\circ}$ meridian passes through the Earth. So as to be consistent with the earlier work of \citet{McConn06}, we have adopted the same values for M31's position angle ($39.8^{\circ}$) and inclination ($77.5^{\circ}$ - \citealt{Vauc1958}). Each object is hence rotated by  $39.8^{\circ}$ about the z-axis to counter the effect of its position angle, and then $77.5^{\circ}$ about the x-axis to account for M31's inclination. A final rotation of $90^{\circ}$ about the z-axis is then necessary to bring $l_{M31} = 0^{\circ}$ into alignment with the direction of Earth (which hence lies at $l_{M31} = 0^{\circ}, b_{M31} = -12.5^{\circ}$). The resulting spherical coordinates for each object in the sample are plotted onto an Aitoff-Hammer projection in Fig. \ref{sat_aitoff}. This same figure also shows the uncertainties in position associated with each object, generated via sampling of the respective distance posterior probability distributions (PPDs) of each object and subsequent conversion of each drawn distance into a three-dimensional position.   

\footnotetext[1]{Defined so as to point north in Equatorial coordinates}

With the satellites' positions determined in cartesian coordinates, it is straight forward to determine the minimum distance of each satellite from a given plane as follows:

 \begin{equation}
    D_{plane} = |ax + by + cz + d|
\label{e_plane_dist}
\end{equation}
where $D_{plane}$ is the distance of a satellite at a point $(x,y,z)$ from a plane whose normal vector is $(a,b,c)$ and is of unit length. For simplicity, we invoke the reasonable requirement that all planes must pass through the center of M31 and so in our case, $d = 0$ and the plane normal vector points out from the center of M31. Hence, in order to find the best-fit or maximum significance plane to a set of satellites, we need simply minimize $D_{plane}$ for the satellites to be fitted. This can be done via a variety of means, some of which are compared in the following section, but perhaps the most robust and the predominant method employed in this study, is that of minimizing the root-mean-square (RMS) of the distances to the fitted satellites. 

In order to measure the asymmetry of the satellite distribution about a given plane, we need only count the number of satellites on one side of the plane. To do this, we can simply remove the absolute value signs from equation \ref{e_plane_dist}, so that the side of the plane on which a satellite lies can be determined by whether $D_{plane}$ is positive or negative. The plane of maximum asymmetry is then taken to be that which divides the sample such that the difference in satellite counts for opposite sides of the plane is greatest. 

Whether we wish to determine the best fit plane through a sample of satellites or the plane of maximum asymmetry, we require a system by which a large number of planes can be tested on the sample so that the goodness of fit (or asymmetry) can be calculated for each. To do this, we define each tested plane by its normal vector or pole $(a,b,c)$ so that Eq. \ref{e_plane_dist} can be applied directly. We then rotate this pole to different orientations around the sky in such a way as to `scan' the whole sphere evenly and at a suitably high resolution. In practice, we need to be able to apply this routine many thousands of times for a large number of samples and so a fast computational time is of the essence. To this end,  for a given sample, our algorithm determines the desired plane following a two step procedure.

Firstly, a low resolution scan of the sphere is made to determine the approximate direction on the sky of the pole to the best-fit plane. Only half the sphere actually needs to be scanned since poles lying on the opposite hemisphere correspond to the identical planes flipped upside down. The low resolution scan tests 2233 different poles across the hemisphere. A near-uniform coverage is achieved by decreasing the number of planes tested in proportion to the cosine of the latitude of the planes' pole. This prohibits what would otherwise be an increased coverage at the higher latitudes of the coordinate system. With the pole to the best-fit plane determined in low-resolution, a high resolution search is then made around the identified coordinates at 10 times the resolution. In this way a pole can effectively be found at any of approximately 250,000 evenly spread locations on the hemisphere.      
 
\subsection{Generating Random Satellite Samples}
\label{ss_Method_rr} 

Whilst we are now equipped to identify best-fit planes to our sample and subsamples thereof, it is necessary to have some means of determining the significance of these planes in an absolute sense. The most intuitive way to do this is to perform the same analysis on a randomly generated sample of equal size. In particular, when we are concerned with all possible combinations of a particular number of satellites that can be produced from the whole sample, we are often dealing with a very large number of subsamples and so it is inevitable that some of these subsets of satellites will exhibit a very high degree of planarity. Identical analysis must therefore be performed on random distributions, to see if there are similar numbers of subsets with equal degrees of planarity.   

For this reason, considerable care was taken to design an algorithm capable of providing a unique random realization of the desired number of satellites whenever it is called. The algorithm makes use of the distance PPD for each satellite, and also takes into account the irregular window function (i.e. useable portion) of the PAndAS survey. Each time a satellite is to be added to the random realization, one of the 27 actual satellites is chosen at random and a distance is drawn from its associated PPD. This distance ($D_{sat}$) is then converted into a three dimensional position $(x,y,z)$ following equation set \ref{e_cart_coord} and this satellite-to-M31 separation vector is then spun around to a new, random location in the M31 sky. Note that for each random realization, a new value of $D_{M31}$ is similarly drawn from the M31 distance PPD.     

Once again, care must be taken in this step to ensure that the whole sphere is given equal weight, otherwise there is a higher likelihood for the artificial satellites to be positioned at high latitude. Again, this is remedied by weighting the likelihood by the cosine of the latitude. 

With the new, random location for the satellite chosen, it is then projected back onto the sky as it would appear from Earth and a check is made to ensure that it does indeed lie within the boundaries of the PAndAS survey area, and outside of the central ellipse ($ 5^\circ$ major axis, $2^\circ$ minor axis - see Fig. 10 (c) of CIL12) where the disk of M31 inhibits reliable measurements. If the satellite does not meet these requirements, it is rejected and the satellite drawing process is repeated until a suitable position is generated. By repeating this process until the desired number of satellites are produced, a new, random comparison sample is generated which gives full account to the constraints on the actual data.   

In order for the random satellite realizations to mimic the actual data most closely, it is necessary that each artificial satellite is represented not by just one point, but rather a string of points reflecting the uncertainty in the Earth-to-Object distance. Hence once acceptable positions for each satellite are drawn as described above, the distance distributions for each object are sampled and projected to their equivalent positions along the line of sight about the initially placed point. For sections \ref{ss_BFP_all_sats}, \ref{ss_asy_all_sats} and \ref{Great_Plane} each artificial satellite's distance distribution is represented by $1000$ points such that each plane-fitting measurement is made for $1000$ possible positions of the object and then the average value of the measurements is taken. The only exception to this number is where the maximum-likelihood approach is used in \S \ref{ss_BFP_all_sats}. Due to the inclusion of a second fitting parameter in this case, only $100$ samples are taken for each satellite. For \S \ref{sat_subsets}, as we are not concerned with comparisons of plane significance between the real sample and the random realizations, it is sufficient to use a single drawn position for each artificial satellite.         

\subsection{A note on Satellite Detection Bias}
\label{ss_Method_sdb} 

\begin{figure}[htbp]
\begin{center}
\includegraphics[width = 0.50\textwidth,angle=0]{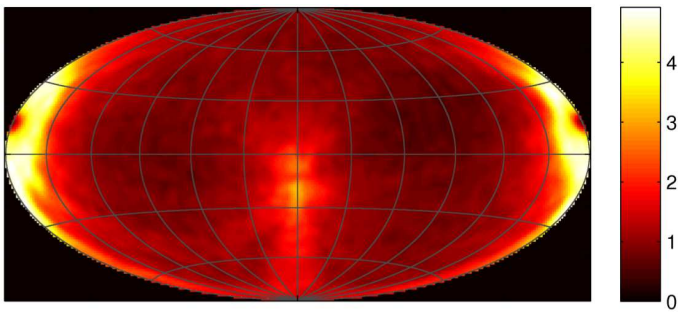}				  
\caption{An Aitoff-Hammer projection illustrating the satellite detection bias resulting from the PAndAS survey boundaries and M31 disk obstruction. Note that this figure utilizes a Gaussian blurring of radius $5^\circ$, as do all of the subsequent pole-density plots.}
\label{survey_mask_effects}
\end{center}

\end{figure}

By employing a similar method to that described above, it is also possible to explore the effect of the PAndAS survey area boundaries on the satellite detection bias as viewed from the center of M31. It is intuitive that more satellites are likely to be detected along the line of sight to Earth, since even satellites at a large distance from M31 will still appear within the survey boundaries if they lie along this line. We can visualize this effect by generating a large number of randomly distributed satellites and plotting them on the M31 sky after first rejecting those satellites that would appear outside the survey area `mask' if viewed from Earth. To do this, one million satellites were drawn from a spherically symmetric halo potential with density falling off as a function of the square of the distance from the halo center. Satellites were hence drawn at distances between $0$ and $700$ kpc from M31 with equal probability. The satellites were then projected onto the M31 tangent plane and those satellites lying outside the survey area or inside the M31 disk obstruction area were excised from the density map. The resulting anisotropy of the satellites on the M31 sky is presented in Fig. \ref{survey_mask_effects}.

 As can be seen from the figure, the probability of detection is indeed higher along a great circle oriented edge-on with respect to the direction of Earth, and perpendicular to the M31 disk ($b_{M31} = 0^{\circ}$). This great circle has its pole/ anti-pole at $l_{M31} = \pm 90^{\circ}, b_{M31} = 0^{\circ}$ and hence we would expect a predisposition toward finding planes of satellites with a pole in this vicinity. We would also expect, though to a lesser extent, to find an excess of satellite planes oriented edge-on with respect to Earth at any inclination. Such planes would have poles lying anywhere on the great circle whose normal is directed toward Earth. The drop in the satellite density at $l_{M31} = 0^{\circ}, b_{M31} = -12.5^{\circ}$ and $l_{M31} = \pm 180^{\circ}, b_{M31} = 12.5^{\circ}$ is a consequence of the hinderance to detection caused by the M31 disk. Due to the increased volume of space covered by the survey at greater distances from Earth, unhindered satellite detection is possible over a larger range of angles on the far side of M31 in comparison to the Earth-ward side.



\section{Results}
\label{s_results}

\subsection{Best Fit Plane to the Entire Satellite Sample}
\label{ss_BFP_all_sats}

In order to find the best-fit plane to the satellite system as a whole, the procedure of \S \ref{ss_Method_pf} is applied to the whole sample of 27 satellites presented in CIL12. The RMS thickness of the sample is used here, as in subsequent sections, as the statistic of planarity; we find it to be a robust measure and it has the convenient property of being computationally inexpensive. Since we are dealing with only one sample in this case, two other measures are also used for comparison. The first calculates the sum of the absolute values of the distances of each of the satellites from the tested plane. The second is essentially a maximum likelihood approach and replaces the plane of zero-thickness with a `Gaussian Plane' such that a satellite's position within the Gaussian determines the plane's goodness-of-fit to that satellite. This second approach requires that different Gaussian widths $\sigma$ be tested for each plane orientation in order to find the width that best matches the satellite distribution. Values between 5kpc and 150 kpc were tested at 5 kpc intervals for each tested plane orientation.  Hence an additional characteristic of the satellite distribution is obtained, but at the expense of a considerably longer computation time.

\begin{figure}[htbp]
\begin{center}
\includegraphics[width = 0.26\textwidth,angle=-90]{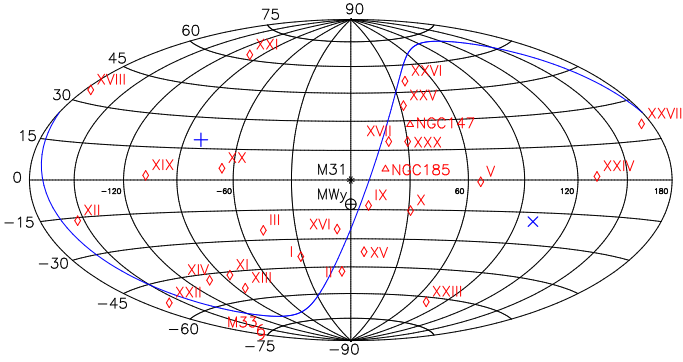}				  
\caption{An Aitoff-Hammer Projection showing the best-fit plane to the satellite system as a whole. The pole and anti-pole of the plane are denoted by `$+$' and `$\times$' symbols respectively. Only the best-fit satellite positions were incorporated into the fit for this figure. The distribution of poles obtainable from other possible realizations of the satellite distribution is presented in Fig. \ref{RMS_distribution}. Note that the plane is near-polar, similar to the preferred plane orientations identified for the Milky Way Satellite System.}
\label{bfp_all_sats}
\end{center}

\end{figure}

For each of the three measures of goodness-of-fit described above, the first step is to find the best-fit plane to the satellite positions with their positions determined from their best-fit distances. When either the RMS or maximum likelihood approach is used, the same best-fit plane is found as $0.153x + 0.932y + 0.329z = 0$ with pole at $(l_{M31}, b_{M31}) = (-80.7^{\circ}, 19.2^{\circ})$. This plane is plotted as a great circle on the M31 sky in Fig. \ref{bfp_all_sats} with the poles of the plane indicated. When the absolute distance sum is used instead, the pole is found farther from the plane of the galaxy, at $(l_{M31}, b_{M31}) = (-74.9^{\circ}, 24.3^{\circ})$. Nevertheless, the polar-plane described by \citet{Koch06} is supported by either measurement, and is reminiscent of the satellite streams identified in the Milky Way satellite system. In light of the detection biases imposed by the PAndAS survey area as illustrated in Fig. \ref{survey_mask_effects}, the result in this case must clearly be treated with suitable caution however. Like \citet{Koch06}, we find little evidence for the Holmberg Effect, with only 3 best-fit satellite positions falling within $30^{\circ}$ of the M31 galactic poles, and only 6 of the $1 \sigma$ error trails from Fig. \ref{sat_aitoff} pass beyond $b_{M31} = \pm 60^{\circ}$.

\begin{figure}[htbp]
\begin{center}
\includegraphics[width = 0.50\textwidth,angle=0]{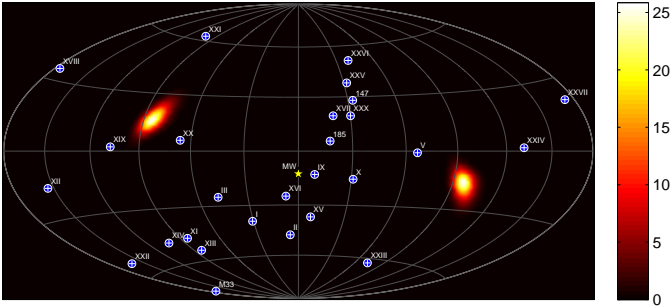}				  
\caption{A pole-density map showing the effective uncertainty in the location of the best-fit plane to the whole satellite sample. The poles of the best-fit planes derived for 200,000 possible realizations of the data are plotted, along with their corresponding anti-poles.}
\label{RMS_distribution}
\end{center}

\end{figure}

 To determine the uncertainty in the plane's goodness-of-fit, we need to repeat the procedure for a large number of realizations of the satellite sample, with the best-fit satellite distances replaced with a distance drawn at random from their respective satellite distance PPDs. A density map of the best-fit plane poles identified from 200,000 such realizations is presented in Fig. \ref{RMS_distribution}. This figure was generated using the distribution RMS as the goodness-of-fit statistic, and contains $71.1 \%$ of all poles within a $5^{\circ}$ radius of the best-fit pole stated above. When the sum of absolute distances is used in place of the RMS, this fraction falls to $68.3 \%$, or to $70.9 \%$ when the maximum likelihood approach is used. It should be noted that the distribution of poles lies in close proximity to the pole of maximum detection bias at $l_{M31} = - 90^{\circ}, b_{M31} = 0^{\circ}$, again suggesting that the detection bias is having a strong influence on the polar orientation of the best-fit plane.  
 
\begin{figure*}[htbp]
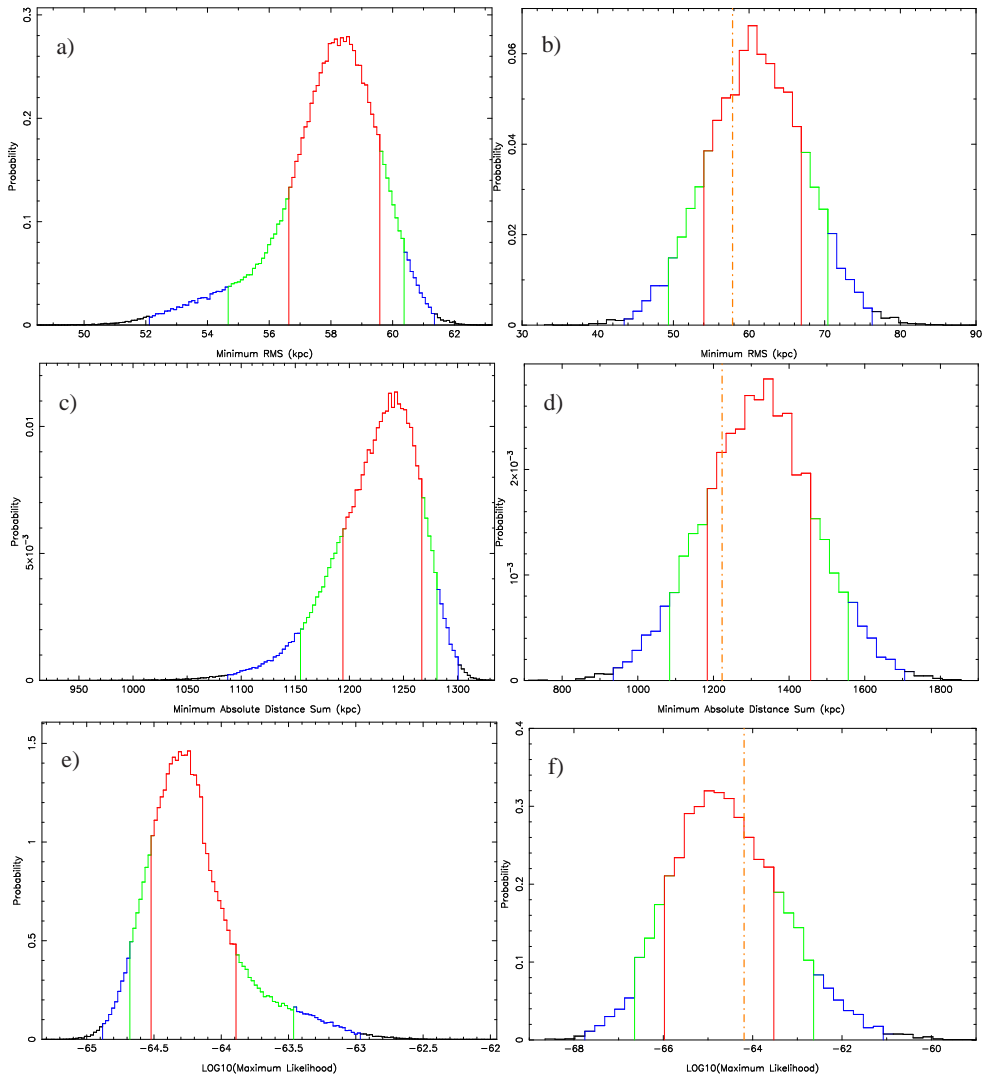

\begin{center} 
$ \begin{array}{c}
\begin{overpic}[width = 0.26\textwidth,angle=-90]{Figures/RMS_err_samp_PPD2.ps}
\put(10,63){\small a)}
\end{overpic}
\begin{overpic}[width = 0.26\textwidth,angle=-90]{Figures/RMS_sig_PPD2.ps}
\put(10,63){\small b)}
\end{overpic}
\\ \begin{overpic}[width = 0.26\textwidth,angle=-90]{Figures/AbVal_err_samp_PPD2.ps}
\put(10,63){\small c)}
\end{overpic}
\begin{overpic}[width = 0.26\textwidth,angle=-90]{Figures/AbVal_sig_PPD2.ps}
\put(10,63){\small d)}
\end{overpic}
\\ \begin{overpic}[width = 0.26\textwidth,angle=-90]{Figures/ML_err_samp_PPD2.ps}
\put(10,63){\small e)}
\end{overpic}
\begin{overpic}[width = 0.26\textwidth,angle=-90]{Figures/ML_sig_PPD2.ps}
\put(10,63){\small f)}
\end{overpic}

\end{array}$
\end{center}
\caption{Probability distributions for the planarity of the entire satellite sample, as determined from three different measures of the plane goodness-of-fit. The left-hand column of figures gives the distribution of the goodness-of-fit statistic as obtained via plane fitting to 200,000 separate samplings of the \emph{real} satellite sample. The right-hand column of figures summarizes the same procedure performed for $1,000$ separate samplings of each of $10,000$ \emph{random} realizations of the satellites (as per \S \ref{ss_Method_rr}). It is important to note that each histogram in this column has been generated by plotting the \emph{average} values from the $10,000$ individual histograms corresponding to each of the random realizations and hence they should only be compared with the \emph{average} of the histograms in the left-hand column. The goodness-of-fit statistic for a) and b) is the distribution RMS; for c) and d) is the absolute distance sum and; for e) and f) is the sum of satellite likelihoods. The average of the histograms in (a), (c) and (e) are shown in (b), (d) and (f) respectively as dashed lines. Red, green and blue lines denote the extent of $1 \sigma$ ($68.2 \%$), $90 \%$ and $99 \%$ credibility intervals respectively.}
\label{BFP_PPDs}

\end{figure*}

 In order to determine whether the goodness-of-fit of the best-fit plane is really physically significant,  similar analysis should be performed on a large number of random realizations of satellites, to see how often distributions of satellites arise with a comparable degree of planarity. Figure \ref{BFP_PPDs} presents probability distributions of the plane significance for possible realizations of the real satellite sample along with average values from random realizations of the satellites (as per \S \ref{ss_Method_rr}), obtained using the three measures of goodness-of-fit stated above.

It is immediately clear from Fig. \ref{BFP_PPDs} that regardless of the choice of the measure of goodness-of-fit, the range of values obtainable from possible realizations of the real satellite positions are similar to the most likely values to be expected from completely random realizations of the satellites. Hence, whilst a prominent plane of satellites comprising roughly half of the sample is suggested in Fig. \ref{sat_aitoff}, it would seem that the sample as a whole is no more planar than  would be expected from a strictly random distribution. Again, this is in keeping with the findings of \citet{Koch06}, and detracts from any physical significance that should be attributed to the plane's polar orientation. 

\begin{figure}[htbp]
\begin{center}
\includegraphics[width = 0.26\textwidth,angle=-90]{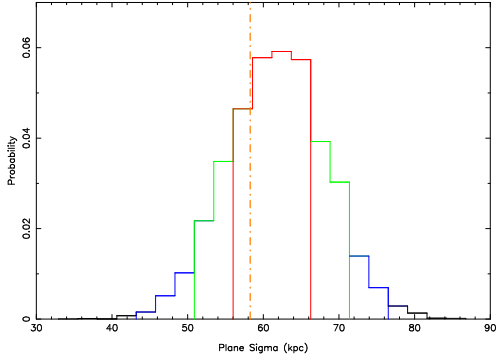}				  
\caption{The probability distribution for the \emph{average} $1 \sigma$ width as determined from 10,000 random distributions of $27$ satellites. This figure is generated from the same run as Fig. \ref{BFP_PPDs} f) and is the result of marginalizing over the plane-orientation model parameters.}
\label{bfp_sigma}
\end{center}

\end{figure}

Further to this finding, the overall width of the `plane' is again in keeping with that expected from a purely random satellite distribution. From fitting the Gaussian Plane to the best-fit satellite positions, a $1 \sigma$ width of $60$ kpc is found to produce the best fit to the data. When the 200,000 PPD-sampled realizations were tested, a $1 \sigma$ of $60$ kpc was found preferential in $66.3 \%$ of cases, with a $1 \sigma$ of $55$ kpc being preferred in $32.7 \%$ of cases. Values of $50$ kpc make up the remaining $1 \%$ almost entirely. The average value for the actual satellite distribution was thus determined as $58.3$ kpc. This value is similar to the most likely width identified from the 10,000 random realizations, as can be seen in Fig. \ref{bfp_sigma}.

\subsection{The Plane of Maximum Asymmetry}
\label{ss_asy_all_sats}

\begin{figure}[htbp]
\begin{center}
\includegraphics[width = 0.26\textwidth,angle=-90]{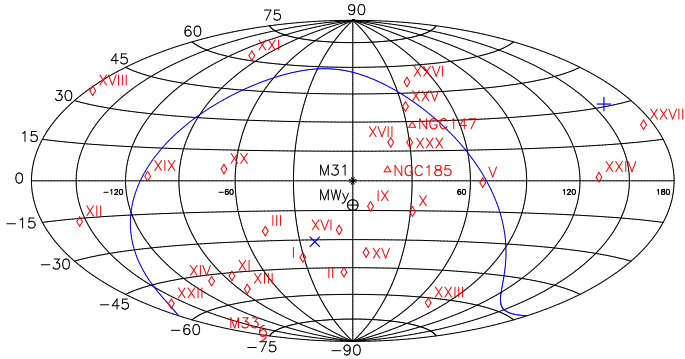}				  
\caption{An Aitoff-Hammer projection showing the plane of maximum asymmetry identified from the full sample of best-fit satellite positions. It divides the distribution such that 21 satellites lie in one hemisphere, but only $6$ in the other. The anti-pole of the maximum asymmetry plane lies just $28.1^{\circ}$ from the Milky Way as viewed from the center of M31.}
\label{bfp_asy}
\end{center}

\end{figure}

To determine the plane of maximum asymmetry and its significance, we employ an identical approach as in the preceding section, but with the goodness-of-fit statistic replaced with a count of the number of satellites on each side of the plane as per \S \ref{ss_Method_pf}. As was suggested by the three-dimensional satellite distribution generated in CIL12, the asymmetry about the M31 tangent plane is close to a maximum, with $19$ satellites on the near-side of the plane but only $8$ on the other when the best-fit satellite positions are assumed. The highest asymmetry plane possible from this same distribution has $21$ satellites on one side and $6$ on the other, with the equation of the plane identified by the algorithm as $-0.797x - 0.315y + 0.515z = 0$. The anti-pole of this plane lies $27.2^{\circ}$ away from the Milky Way at $(l_{M31}, b_{M31}) = (-21.6^{\circ}, -31.0^{\circ})$. This plane is plotted as a great circle on the M31 sky in Fig. \ref{bfp_asy}.

\begin{figure}[htbp]
\begin{center}
\includegraphics[width = 0.52\textwidth,angle=0]{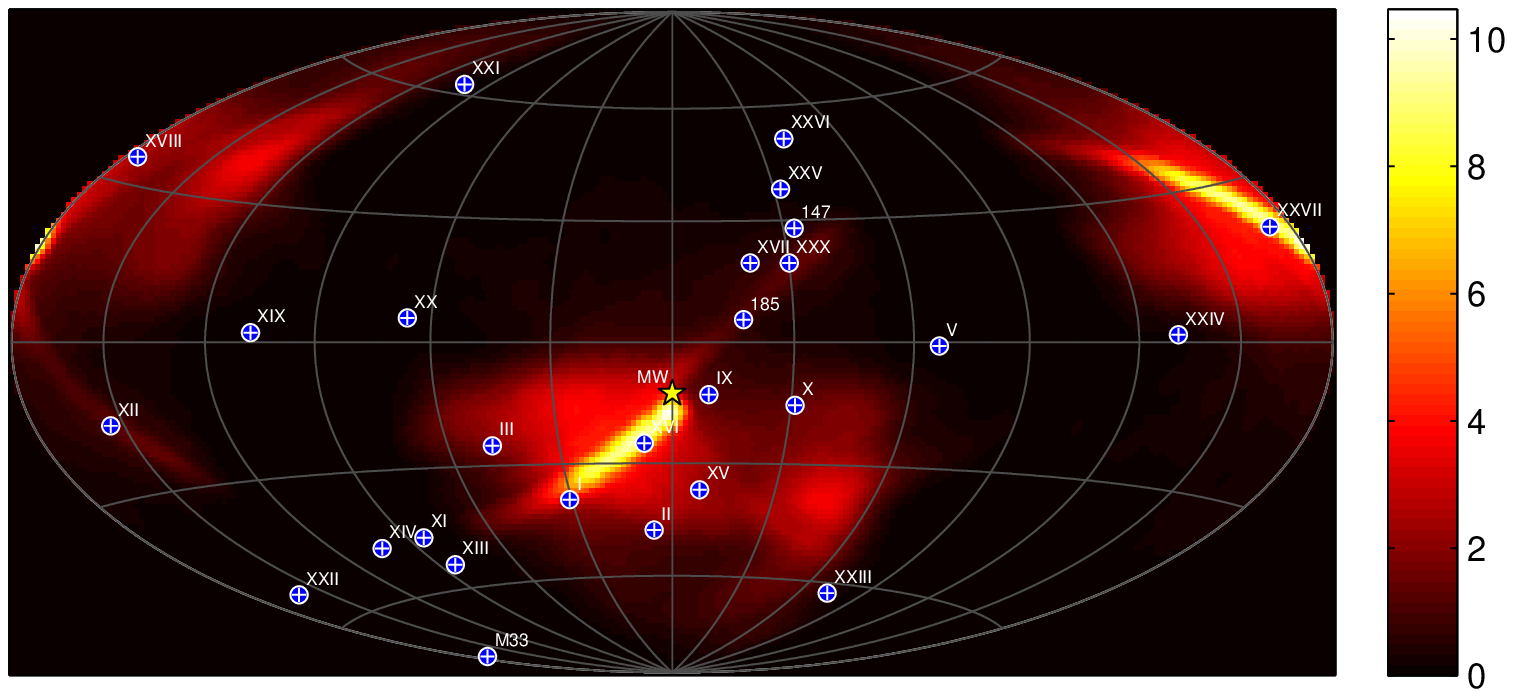}				  
\caption{A pole-density map showing the effective uncertainty in the location of the maximum asymmetry plane to the whole satellite sample. The poles of the maximum asymmetry planes derived for 200,000 possible realizations of the data are plotted, along with their corresponding anti-poles. The elongated distributions that run through the pole and anti-pole determined from the best-fit distribution (see Fig. \ref{bfp_asy}) arise due to the orientation of the uncertainty trails of the individual satellite positions, as presented in Fig. \ref{sat_aitoff}. Note that the probability of the anti-pole of the asymmetry lying within a couple of degrees of the direction of the Milky Way is close to a maximum.}
\label{asy_distribution}
\end{center}

\end{figure}

\begin{figure*}[htbp]
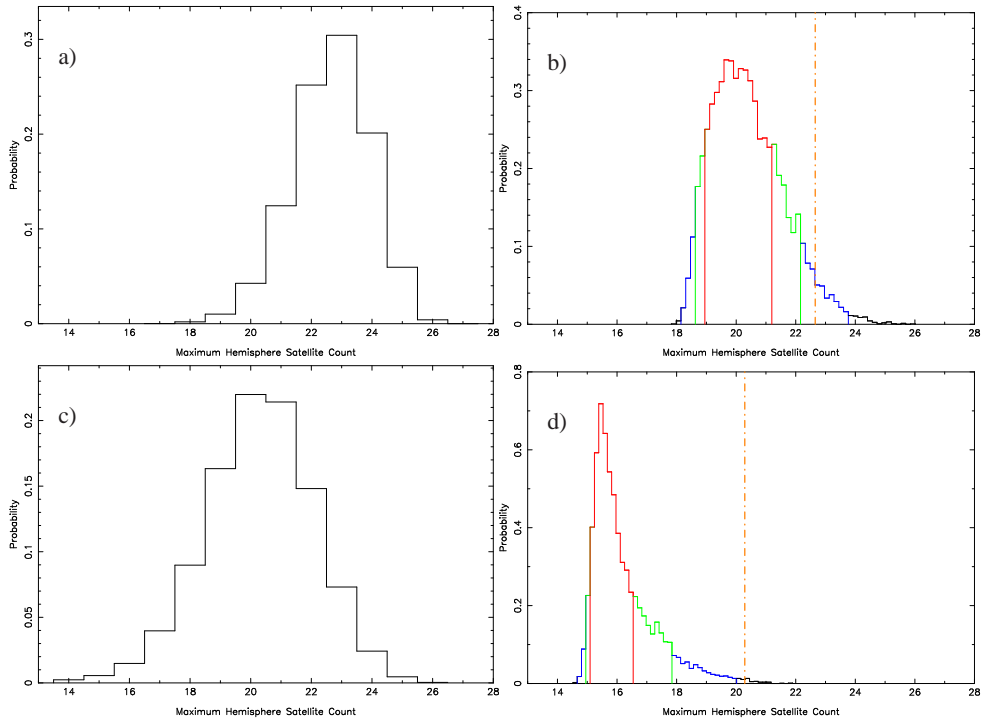

\begin{center} 
$ \begin{array}{c}
\begin{overpic}[width = 0.26\textwidth,angle=-90]{Figures/err_samp_PPD2.ps}
\put(10,60){\small a)}
\end{overpic}
\begin{overpic}[width = 0.26\textwidth,angle=-90]{Figures/asymm_PPD2.ps}
\put(10,60){\small b)}
\end{overpic}
\\ \begin{overpic}[width = 0.26\textwidth,angle=-90]{Figures/err_samp_FP_PPD2.ps}
\put(10,60){\small c)}
\end{overpic}
\begin{overpic}[width = 0.26\textwidth,angle=-90]{Figures/asymm_FP_PPD2.ps}
\put(10,60){\small d)}
\end{overpic}

\end{array}$
\end{center}
\caption{Asymmetry probability distributions. The top two histograms plot probability distributions for the greatest number of satellites that can be found in one hemisphere, as generated from (a) 200,000 samplings of satellite positions possible from the data and; (b) the \emph{average} of $1000$ samplings from each of 10,000 random realizations of the satellites generated as per \S \ref{ss_Method_rr}. Figures (c) and (d) give the equivalent distributions when the maximum asymmetry plane is replaced with the fixed M31 tangent plane. As for Fig. \ref{BFP_PPDs}, the histograms in the right-hand column should only be compared with the average of the corresponding histogram in the left column. The average value of the histograms of (a) and (c) are shown in (b) and (d) respectively as a dashed line.}
\label{Asy_PPDs}

\end{figure*}

When $200,000$ realizations of the satellite sample are generated using the satellite's respective distance probability distributions, the most likely asymmetry of the sample is actually found to be greater than this, with $23$ satellites on one side and only $4$ on the other. Such a scenario is more than twice as likely as the $21:6$ scenario. In one realization, a plane was identified which could divide the sample such that all 27 satellites lay in a single hemisphere, while an asymmetry of $26:1$ was found possible for $815$ $(0.4 \%)$ of the realizations. The distribution of maximum-asymmetry poles on the sky, as determined from realizations of possible satellite positions, is illustrated in Fig. \ref {asy_distribution}, whilst Fig. \ref{Asy_PPDs} $(a)$ plots the probability distribution for the greatest number of satellites that can be found in one hemisphere for a given realization of the observed satellite sample. The average value of this distribution is $22.7$ (shown as a dashed line in Fig. \ref{Asy_PPDs} $(b)$), a value which is equalled or exceeded for $422$ out of the $10,000$ random realizations represented in Fig. \ref{Asy_PPDs} $(b)$. A maximum asymmetry ratio of $21:6$, as was observed for the best-fit satellite distribution plotted in Fig. \ref{bfp_asy}, is more common however, falling inside the $1 \sigma$ credibility interval.

What is particularly striking about the satellite distribution however, is the orientation of the asymmetry, with the majority of satellites lying on the near-side of the M31 tangent plane. From Fig. \ref{Asy_PPDs} (c), it is clear that the effect of the distance uncertainties lying along the line of sight is to create quite a broad distribution in the level of asymmetry about the tangent plane, though the average is markedly high at $20.3$. To investigate the likelihood of this scenario arising from a random satellite distribution, we measure the average number of satellites on either side of the M31 tangent plane for each of $10,000$ random realizations as per \S \ref{ss_Method_rr}. The results are illustrated in Fig. \ref{Asy_PPDs} $d)$. The observed profile is more-or-less as expected, with a maximum probability close to the minimum possible asymmetry at $14$ and then a rapid fall off toward higher asymmetries. It is therefore clear that the distance uncertainties lying along the line of sight have no significant bearing on the orientation of the asymmetry. Yet the observed degree of asymmetry about the M31 tangent plane is equalled or exceeded in only $46$ of the $10,000$ random satellite realizations and hence is very significant. The possibility that this asymmetry may be a consequence of data incompleteness is currently being examined more closely (see \citealt{Martin13}), although it seems very unlikely. The high degree of asymmetry is still observed even when only the brightest satellites are considered. Furthermore, the data incompleteness appears to be dominated by the boundaries of the PAndAS survey area and obstructed regions which are already taken into account by our analysis. Indeed, one would expect more satellites to be observed on the far side of the M31 tangent plane on account of the increased volume of space covered by the survey at greater distances, an effect clearly visible in Fig. \ref{survey_mask_effects}. 

\subsection{Subsets of Satellites}
\label{sat_subsets}

It is perhaps not surprising that the satellite system of M31, when treated as a whole, is no more planar than one would expect from a random sample of comparable size. Indeed, a similar result was noted for the M31 system by \citet{Koch06}. The existence of outliers in our satellite sample was already clear from Fig. \ref{sat_aitoff} and furthermore, if multiple planes of differing orientation are present as has been suggested for both the Milky Way's satellite system (e.g. \citealt{LB1982}; \citealt{Paw12B}) and the M31 system \citep{McConn06}, then the goodness of fit of the best-fit plane to the entire distribution is of little consequence. For this reason, we now concentrate our analysis on subsets or \emph{combinations} of satellites. Specifically, we perform a pole-count analysis by determining the pole of the best-fit plane to every possible satellite combination of a particular size that can be drawn from the entire sample.

A pole-count analysis is an excellent way of mapping the degree of prominence of various planes that exist within the distribution as a whole, whatever their orientation may be. The choice of combination size is not trivial however. The number of combinations $s$ of a particular number of satellites $k$ that can be drawn from the entire sample of $n$ satellites can be determined as follows:

 \begin{equation}
    s = \frac{n!}{k!(n-k)!}
\label{combinations}
\end{equation}
For reasons that shall be discussed shortly, we will effectively be working with a sample of $25$ satellite positions. It is clear from this equation however that with $25$ satellites forming the entire sample, the total number of combinations that can be drawn may be very large, depending on the number of satellites forming the combinations.  For instance, if $n = 25$ and $k = 13$, there are over $5.2$ million possible combinations that can be drawn. Additionally, if we are to properly account for the uncertainties in the satellite positions, it will be necessary to sample from the distance distributions of each satellite a large number of times for every combination. Given that we must test every possible plane orientation (as per \S \ref{ss_Method_pf}) for every rendition of every combination, the computation times can become impracticable. It is therefore necessary to limit our combination sizes as much as possible. We note however, that the final pole-plot distribution showing the poles of the best-fit planes to each combination, is not so dependent on the combination size as might at first be thought. 

With all the planes tested as per \S \ref{ss_Method_pf} having to pass through the center of M31, the minimum number of satellites that can not be fitted exactly is $3$. This is therefore the smallest combination size we consider. There are $2,300$ combinations of $3$ satellites that can be drawn from the full sample of $25$ satellites. If we increase the combination size considerably to $7$ satellites, there are $480,700$ satellite combinations that can be drawn. Due to an excessive number of combinations beyond this point, this is the largest combination size we consider. But it is critical to note that even if we produce our pole-plot map from combinations of only $3$ satellites we \emph{do not} exclusively find planes consisting of $3$ satellites. If a plane of $7$ satellites exists for instance, then by Eq. \ref{combinations}, such a plane will produce $35$ poles at the same location on the pole plot, where a plane consisting of only $3$ satellites would contribute only one pole. Conversely if we take combinations of $7$ satellites, despite the larger number of possible combinations in total, we become less sensitive to planes made up of less than $7$ satellites. So in a sense, the combination size we choose depends on the satellite planes we wish to be most sensitive to. In practice, we have found that the smaller combination sizes of $3$ and $4$ satellites are particularly useful for identifying the lowest RMS planes congregating around the band of satellites visible in Fig. \ref{sat_aitoff}. The larger combination sizes of $5$, $6$ and $7$ satellites gradually shift toward finding planes closer to the best-fit plane to the entire satellite sample illustrated in Fig. \ref{bfp_all_sats}.  

Noting these points, we proceed as follows. First, the number of satellites per combination $k$ is chosen ($3 \le k \le 7$) and then for each combination, distances are drawn for each of the satellites from their respective posterior distance distributions as provided in CIL12. To give a satisfactory representation of the form of the distributions, each combination is sampled $100$ times. As such, each satellite combination contributes not $1$ pole to the pole density map for the chosen combination size but $100$, with the spread of poles relating the possible orientations of the best-fit plane to the combination, given the error in the individual satellite positions. The contribution of each pole to the density map is also weighted by dividing it by the RMS of the best-fit plane it represents. Thus each pole does not contribute $1$ count, but rather some fraction, depending on how good a fit the plane it represents is to the satellites in the combination. This fraction is also further divided by $100$, since it represents only $1 \%$ of the samples for the combination, as just discussed.  

As stated above, it should also be noted that we effectively limit the total number of satellites in our sample to $25$ for all analysis in this subsection. This is to account for the bound group of satellites consisting of NGC147, NGC185 and And XXX (henceforth the NGC147 group - see \citealt{Irwin13}). Since we suspect that these satellites orbit M31 as a group and since they all lie along the apparent plane identified in Fig. \ref{sat_aitoff}, it is preferable to treat the group as a single object when we are not concerned with measurements of the significance of particular planes. To do this, we take the luminosity weighted centre as an approximation for the center of mass, and treat this determined position as though it were the location of a single satellite. To calculate the luminosity weighted center, we can ignore the contribution from And XXX since it is negligible compared with the contributions of the two dwarf ellipticals. From the Third Reference Catalogue of Bright Galaxies \citep{RC3}, NGC185 is $0.2$ magnitudes brighter than NGC147 in the V-band. Each time the NGC147 group is chosen as one of the `satellites' for a combination, distances to each of NGC147 and NGC185 are sampled from their respective distributions and the luminosity weighted center of the group is determined. As for any other combination, this position, along with all other satellites in the combination, is sampled $100$ times. 

\begin{figure*}[htbp]
\begin{center} 
$ \begin{array}{c}
\begin{overpic}[width = 0.25\textwidth,angle=-90]{Figures/sat_combo_3_sats.ps}
\put(-5,45){\small 3 Satellites}
\end{overpic}
\begin{overpic}[width = 0.52\textwidth]{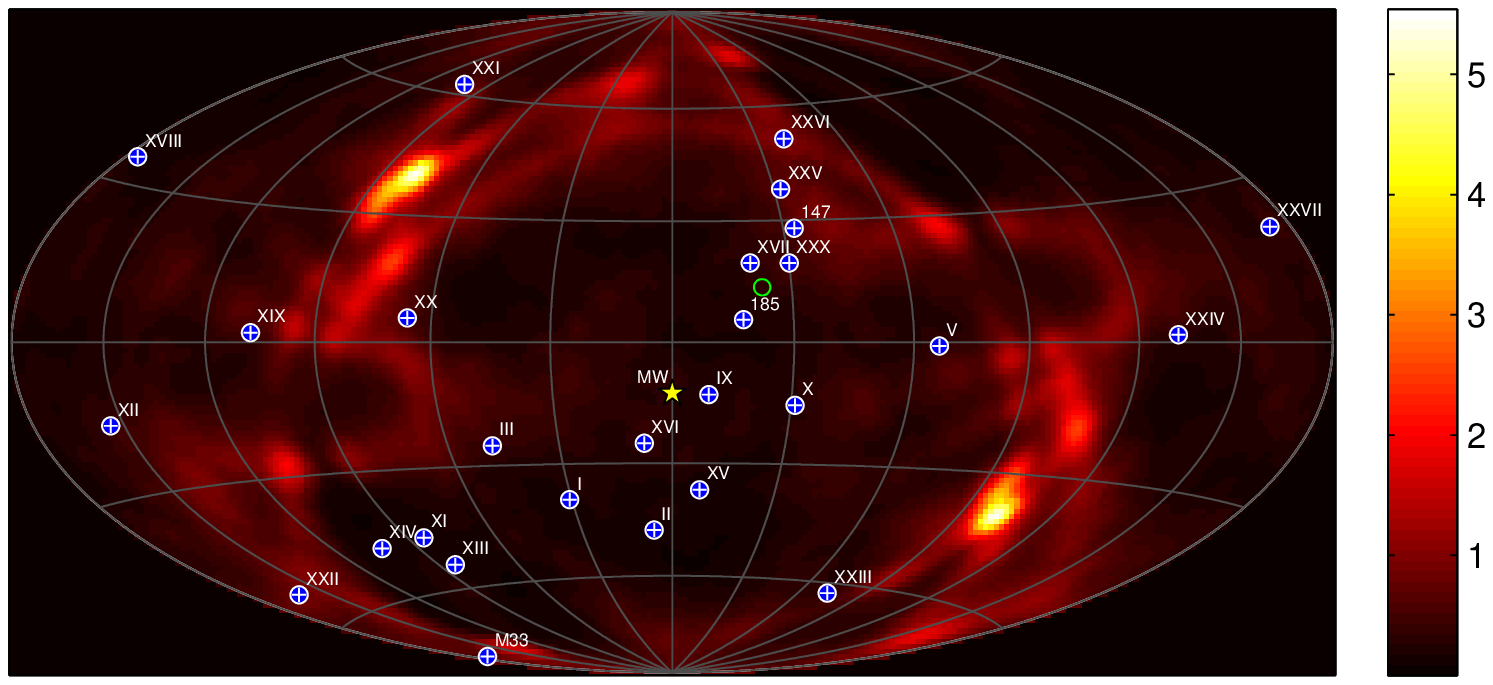}
\end{overpic}
\\ \begin{overpic}[width = 0.25\textwidth,angle=-90]{Figures/sat_combo_4_sats.ps}
\put(-5,45){\small 4 Satellites}
\end{overpic}
\begin{overpic}[width = 0.52\textwidth]{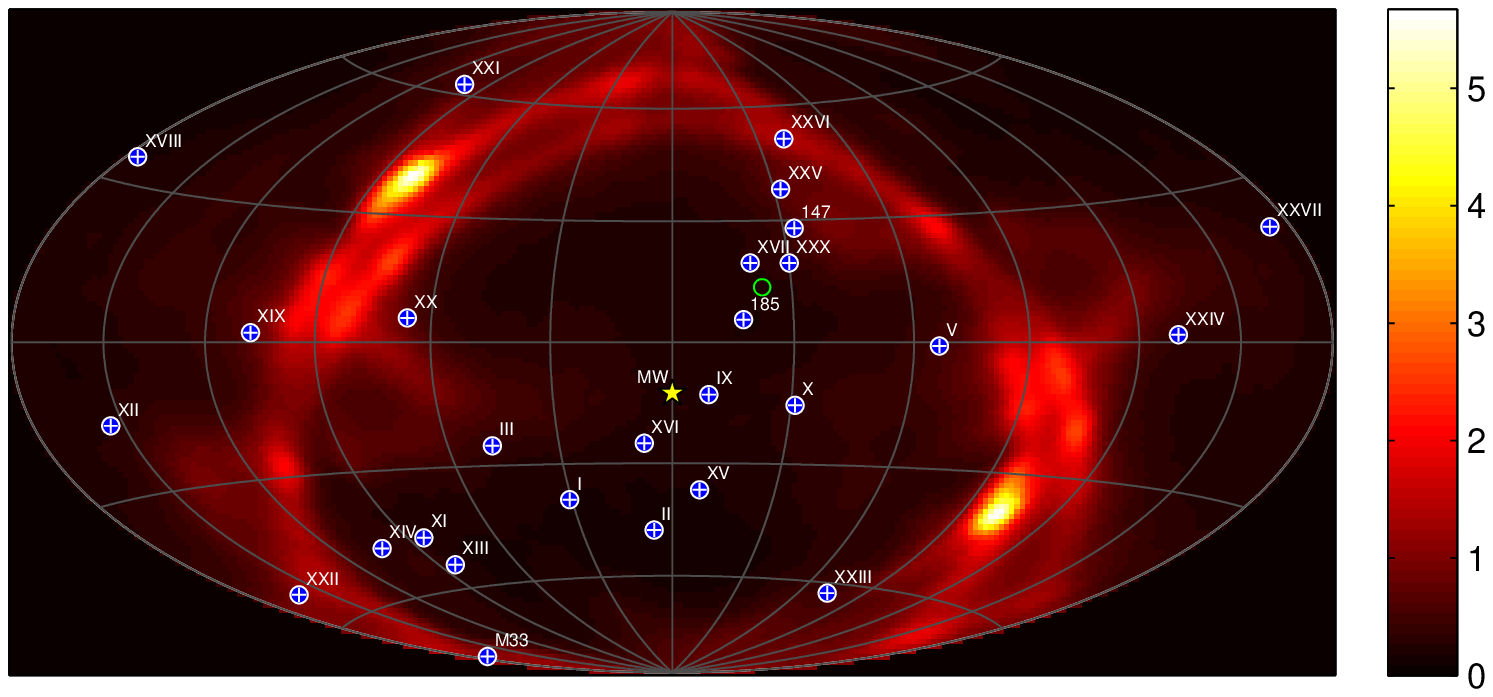}
\end{overpic}
\\ \begin{overpic}[width = 0.25\textwidth,angle=-90]{Figures/sat_combo_5_sats.ps}
\put(-5,45){\small 5 Satellites}
\end{overpic}
\begin{overpic}[width = 0.52\textwidth]{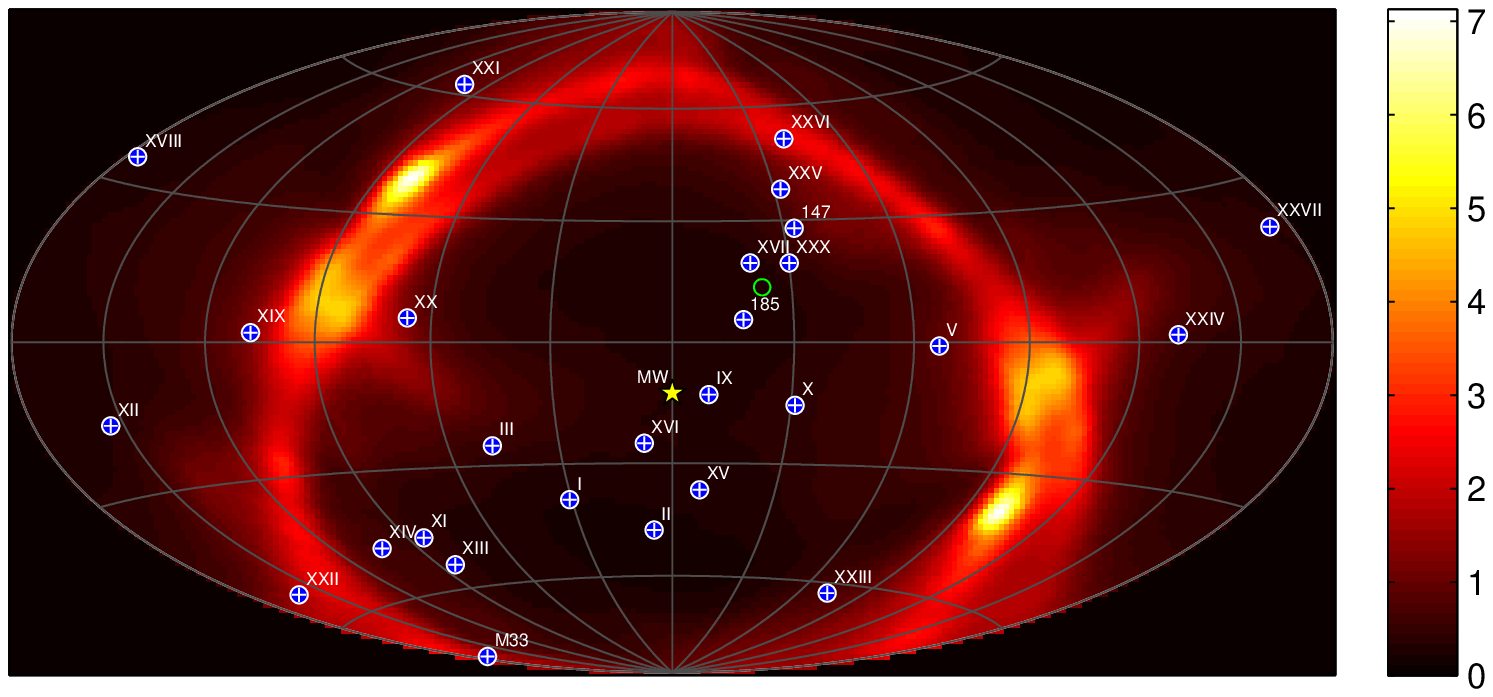}
\end{overpic}
\\ \begin{overpic}[width = 0.25\textwidth,angle=-90]{Figures/sat_combo_6_sats.ps}
\put(-5,45){\small 6 Satellites}
\end{overpic}
\begin{overpic}[width = 0.52\textwidth]{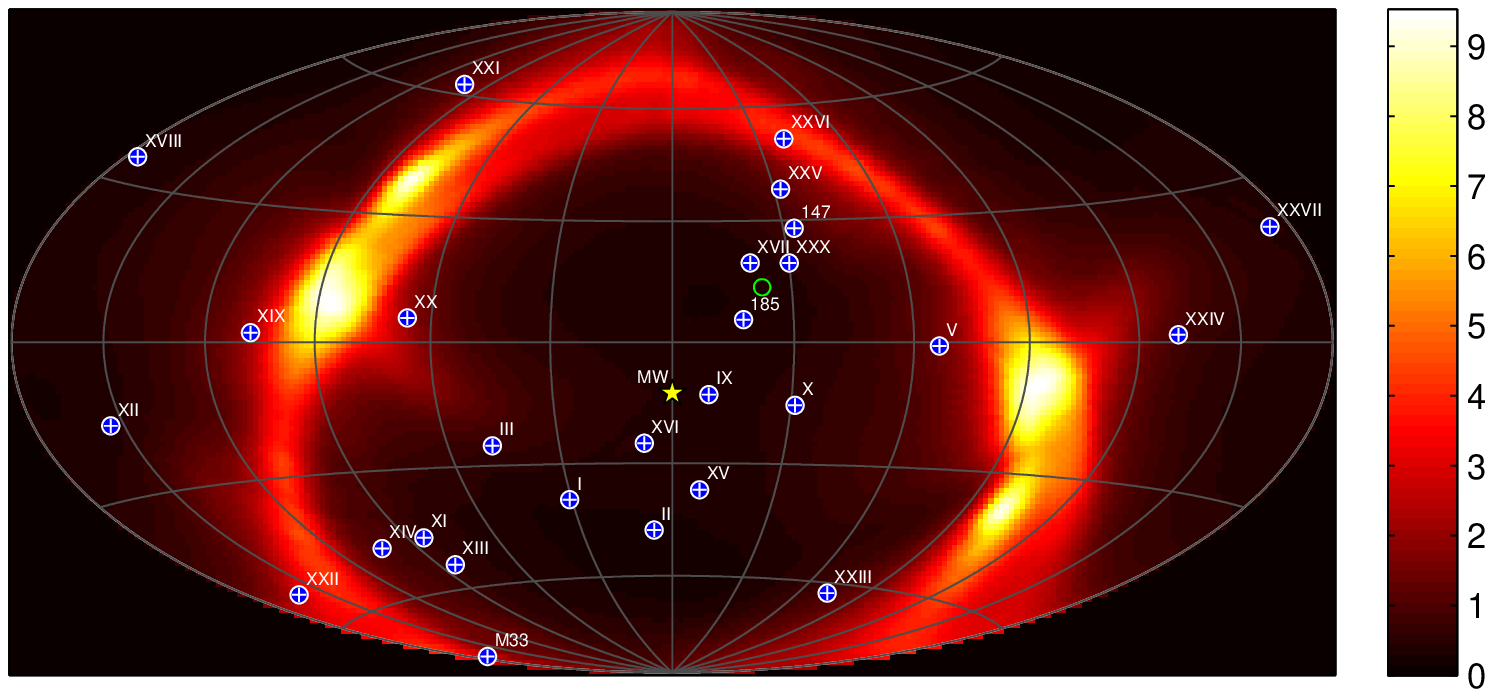}
\end{overpic}
\\ \begin{overpic}[width = 0.25\textwidth,angle=-90]{Figures/sat_combo_7_sats.ps}
\put(-5,45){\small 7 Satellites}
\end{overpic}
\begin{overpic}[width = 0.52\textwidth]{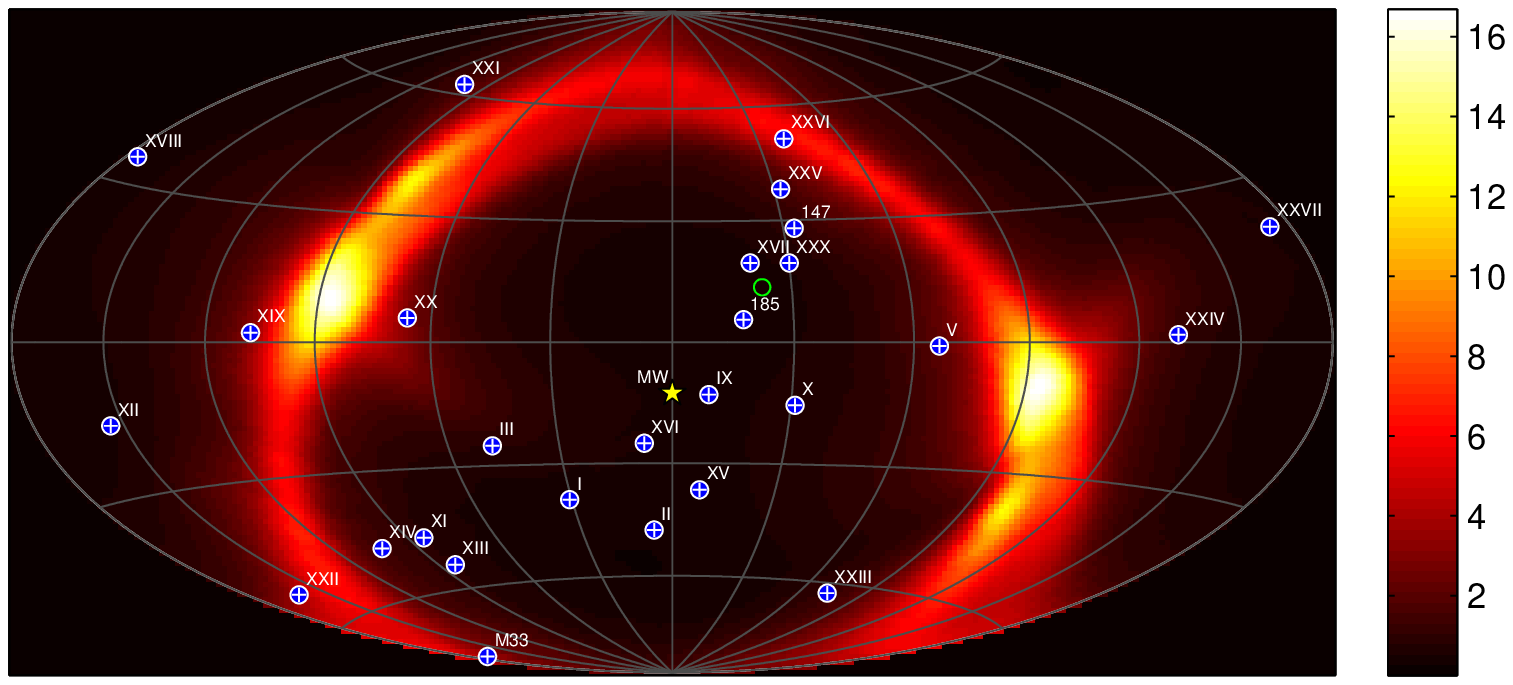}
\end{overpic}

\end{array}$
\end{center}
\caption{Best fit planes and pole density maps for combinations of 3 through 7 satellites. The left-hand column shows the best-fit plane through the combination of satellites that can be fit with the lowest RMS. Satellites included in the best-fit combination are colored red. The centre of the NGC147 group is marked with a circle, and lies on the best-fit plane in every case. The three members of this group are colored orange. Only the best-fit satellite positions are considered for these plots. The right-hand column shows the corresponding pole density plot for the poles of \emph{all} satellite combinations. These plots have been weighted by the RMS of each pole and fully account for the uncertainty in the satellite positions.}
\label{Actual_Pole_Plots}

\end{figure*}

The results of applying the above procedure to all combinations of $3$, $4$, $5$, $6$ and $7$ satellites that can be drawn from the total sample is presented in Fig. \ref{Actual_Pole_Plots}. The left-hand column shows the fit to the most planar combination determined from the best-fit positions whilst the right-hand column shows the corresponding pole density plots for all combinations of that particular number of satellites, based on $100$ samples of each combination as per the discussion above. It is noteworthy that the best-fit planes to the most planar combinations are almost identical in every case, except for that of the $3$ satellite combinations, where the RMS values are so small for so many combinations as to make this result not particularly important. It should also be noted that these best-fit planes trace out the same approximate great circle as the prominent plane indicated in Fig. \ref{sat_aitoff}, a result that shall be investigated a little later in \S \ref{Great_Plane}. It is particularly interesting that the pole shared by each of these planes, located at $ l_{M31} \approx -80^{\circ}, b_{M31} \approx 40^{\circ}$ corresponds to a pole count maximum in each of the pole plots. This indicates that many of the satellite combinations are aligned along this plane, hence further suggesting that the plane applies to more satellites than the combination sizes tested here. The other, lower latitude principle maximum in the pole plots is that corresponding approximately to the best fit to \emph{all} the satellites and hence it grows more prominent in the plots made from larger combination sizes as discussed earlier. 

\begin{figure*}[htbp]
\begin{center} 
$ \begin{array}{c}
\begin{overpic}[width = 0.76\textwidth,]{Figures/pole_density_5_sats.eps}
\end{overpic}
\\ \begin{overpic}[width = 0.50\textwidth]{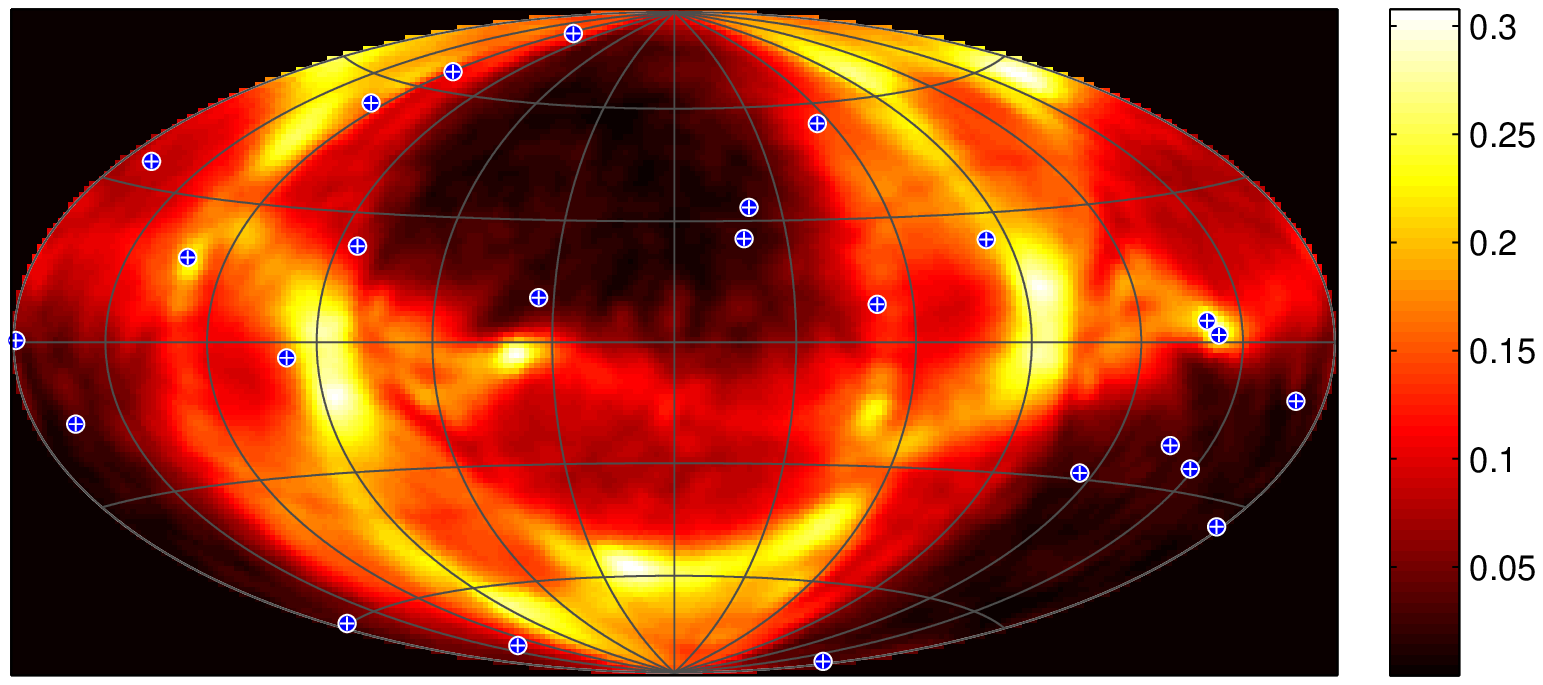}
\end{overpic}
\begin{overpic}[width = 0.50\textwidth]{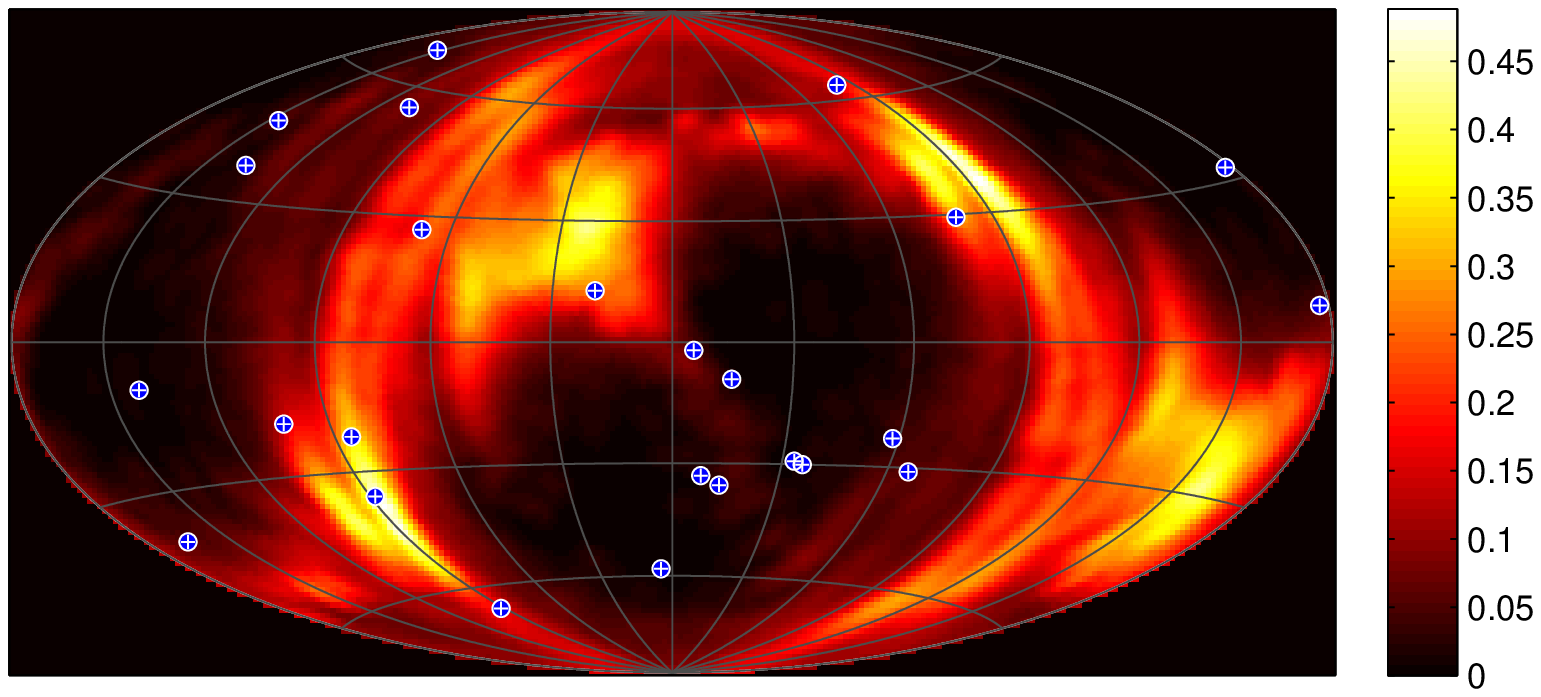}
\end{overpic}
\\ \begin{overpic}[width = 0.50\textwidth]{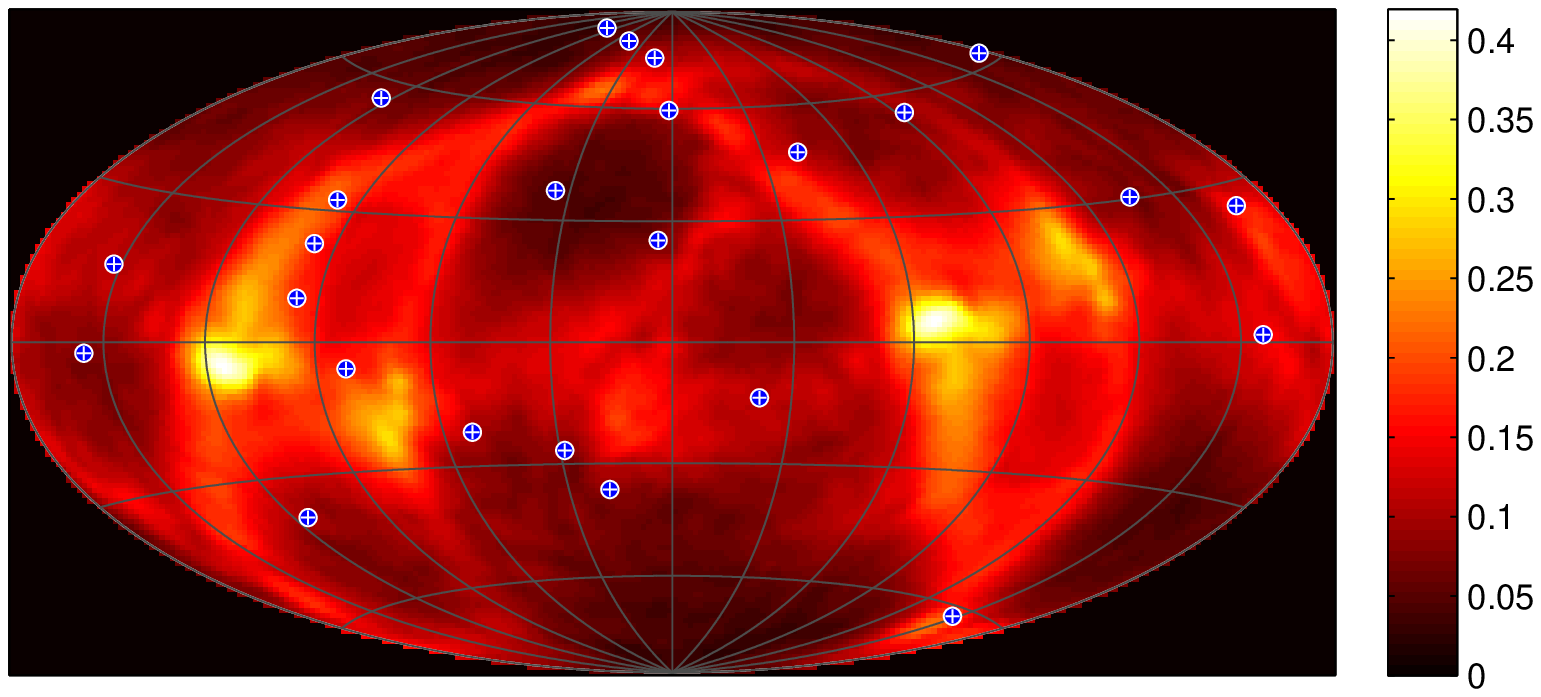}
\end{overpic}
\begin{overpic}[width = 0.50\textwidth]{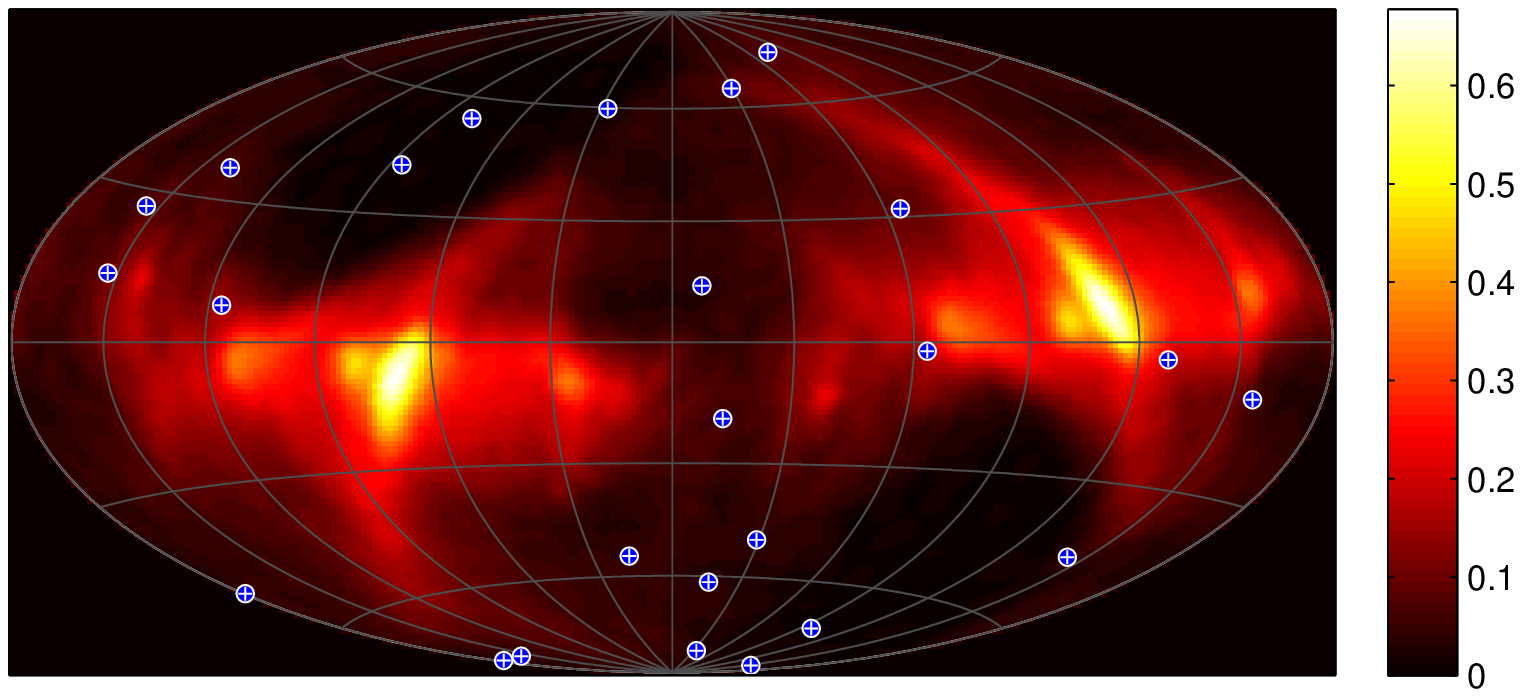}
\end{overpic}
\\ \begin{overpic}[width = 0.50\textwidth]{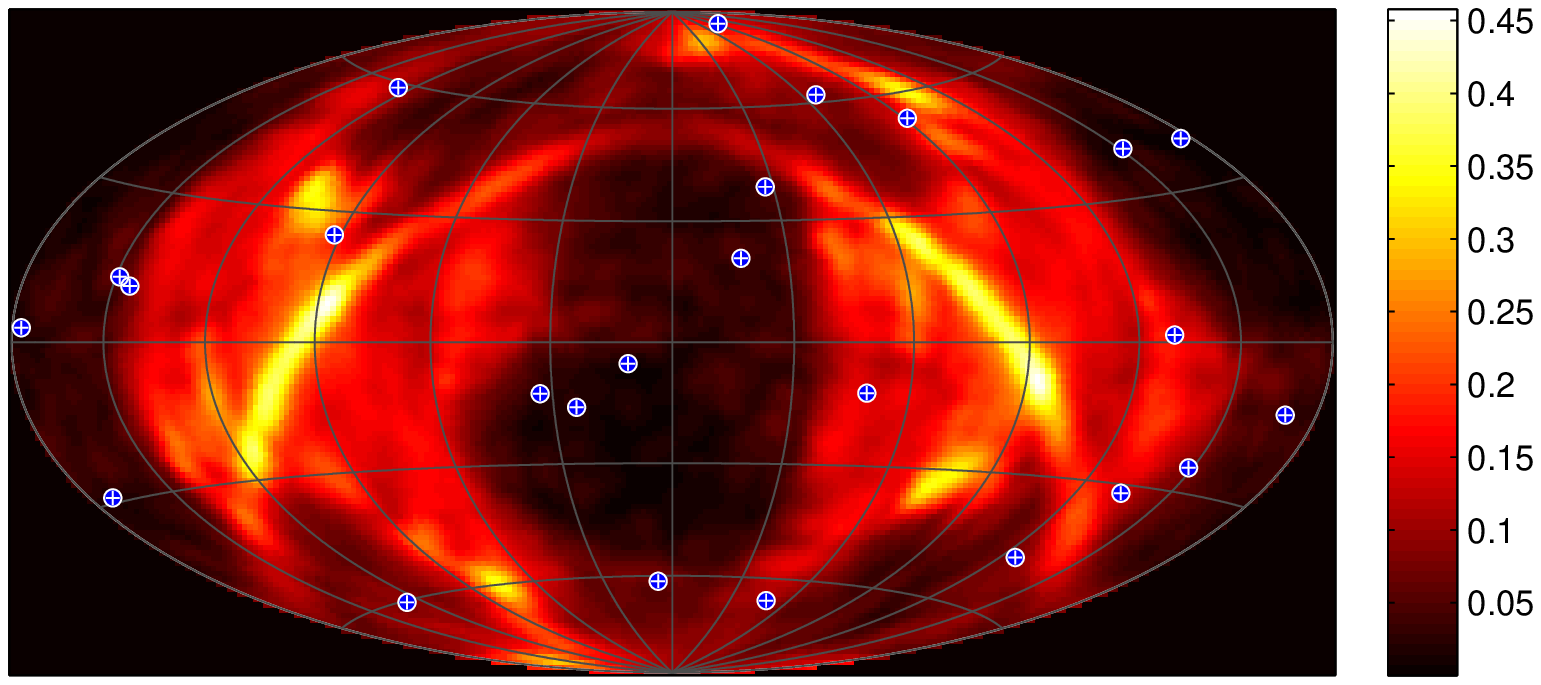}
\end{overpic}
\begin{overpic}[width = 0.50\textwidth]{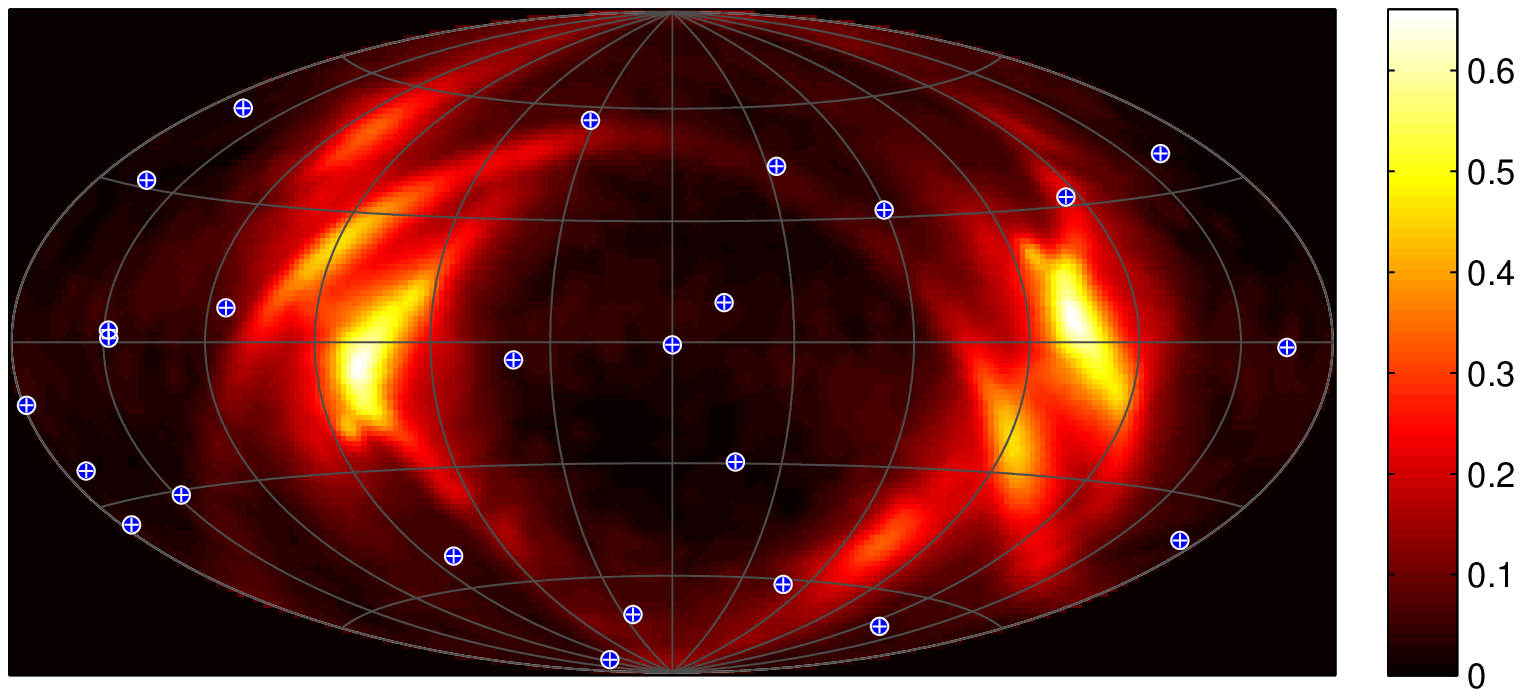}
\end{overpic}
\\ \begin{overpic}[width = 0.50\textwidth]{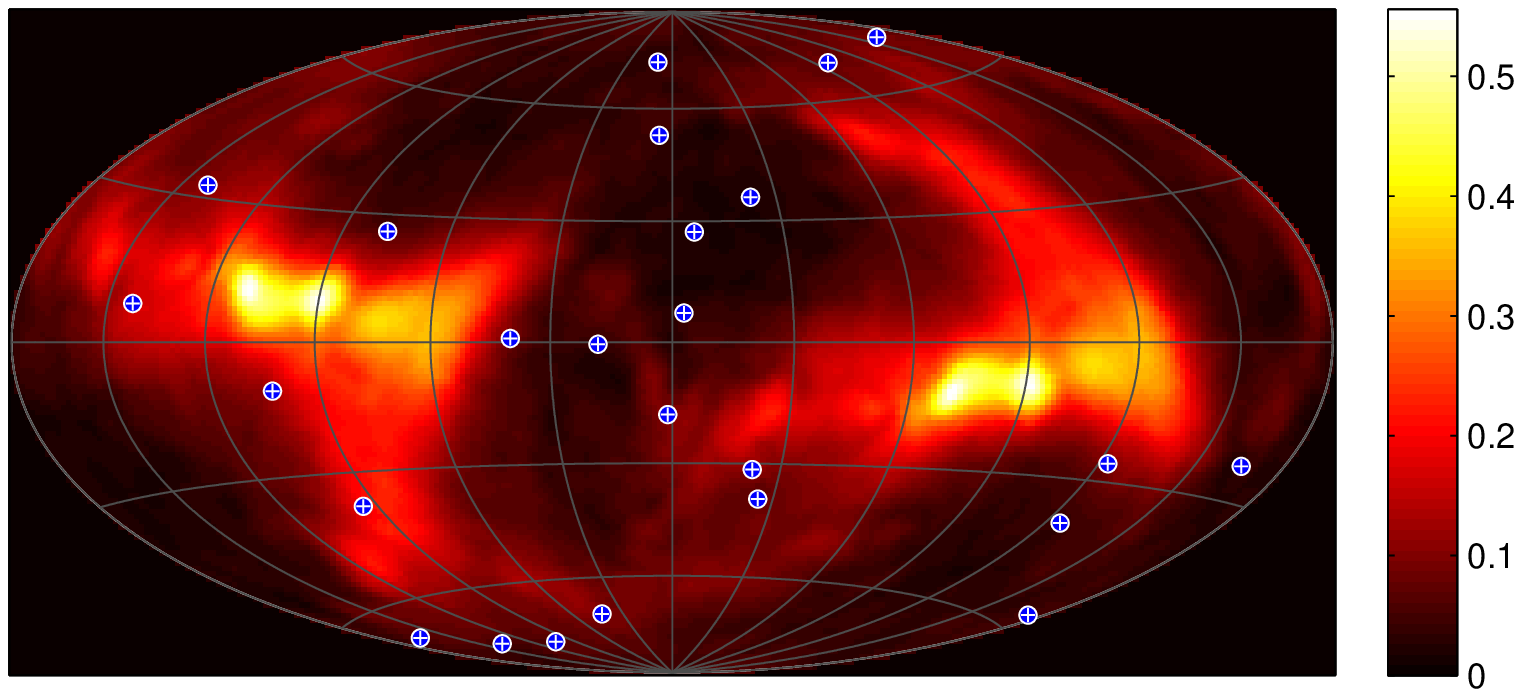}
\end{overpic}
\begin{overpic}[width = 0.50\textwidth]{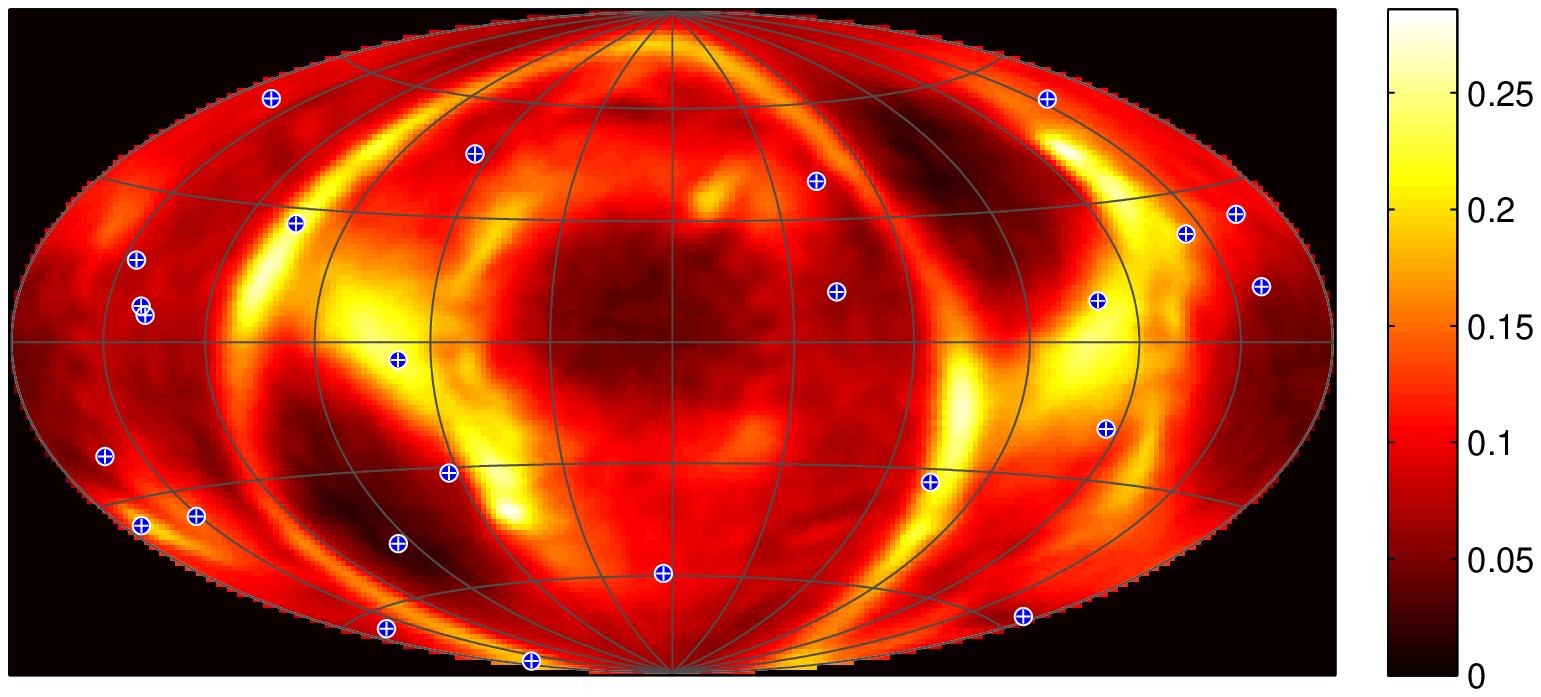}
\end{overpic}

\end{array}$
\end{center}
\caption{Pole density maps for $8$ random realizations of 25 satellites. The maps plot the poles for the best-fit planes to all combinations of $5$ satellites. The contribution of each pole is weighted by the RMS of the plane it represents. The map resulting from all combinations of 5 satellites drawn from the real data is shown again at the top for comparison. Great circle overdensities are evident to varying degrees in the plots and are a result of the satellite detection bias induced by the finite PAndAS survey area. Note that the density scale for the random realizations is much smaller than for the real data on account of the many realizations utilized for each combination of satellites from the real data.}
\label{Random_Reals}

\end{figure*}

Besides the pole count maxima that are strongly indicative of a highly planar subset of satellites, the other principle feature of the pole plots in Fig. \ref{Actual_Pole_Plots} is the great circle along which the pole count density is highest. This great circle is very prominent but great caution must be exercised in attributing any significance to it. It is centered on the Milky Way indicating that the constituent poles result from a majority of satellites lying along the Earth to M31 line of site. But this reflects the anisotropy predicted from Fig. \ref{survey_mask_effects}, the result of the bias incurred by the finite area of the PAndAS survey. Hence it would seem that the progenitor of this prominent great circle is not physical but rather the result of selection effects. To further investigate the significance of the patterns observed in the pole plots, $1000$ random realizations of $25$ satellites were generated as per \S \ref{ss_Method_rr}, and a similar pole count analysis performed on each of them. Specifically, the pole density distribution resulting from the best fit planes to all combinations of $5$ satellites was generated for each of them. The resulting pole plots for $8$ of the $1000$ random realizations (chosen at random) are presented in Fig. \ref{Random_Reals} along with an enlarged version of the equivalent plot from Fig. \ref{Actual_Pole_Plots} generated from the real distribution. A bias toward a similar high-density great circle is indeed observed in these plots, but the plot generated from the actual data features a conspicuously narrower great circle, and a much more constrained distribution in general. This appears to be primarily the result of the large fraction of satellites that lie along the prominent plane that is repeatedly identified and plotted in the left-hand column of Fig. \ref{Actual_Pole_Plots}. It should also be noted that this plane, whilst being oriented perfectly edge-on with respect to the Earth, contains a significant fraction of satellites lying well outside the region of the M31 sky where the detection bias is large, and hence it is unlikely that its prominence is due to our observational constraints.

\begin{figure}[htbp]
\begin{center} 

\includegraphics[width = 0.34\textwidth,angle=-90]{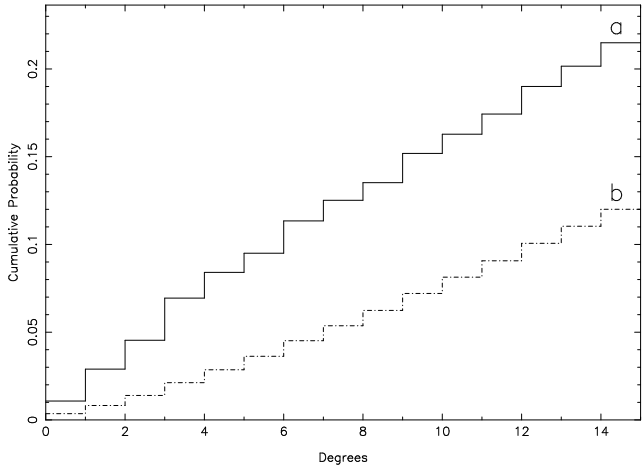}

\end{center}
\caption{Radial density profiles showing the percentage of all poles lying within $n$ degrees of the densest point in the pole count distributions (the principal maximum) for a) the actual satellite distribution and b) the average of 1000 random satellite distributions. The profile for the actual satellite distribution is generated from the same pole distribution as illustrated for $5$ satellites in Fig. \ref{Actual_Pole_Plots} and at the top of Fig. \ref{Random_Reals}. Note that the relative linearity of (b) compared with (a) is simply a result of the averaging of a large number of individual profiles undertaken to produce the former.}
\label{Pole_Radial_Densities}

\end{figure}

Figure \ref{Pole_Radial_Densities} provides for a comparison between the concentration of poles around the principle maximum in the pole distributions of the actual satellite distribution and the average of the $1000$ random satellite distributions. From line (a) in Fig. \ref{Pole_Radial_Densities} we see that $21.5 \%$ of all combinations of the actual satellite positions are fitted by a best-fit plane with pole within $15^{\circ}$ of the principal maximum (located at $l_{M31} = -78.7^{\circ}, b_{M31} = 38.4^{\circ}$). This is in stark contrast to the $12.0 \%$ that lie within $15^{\circ}$ of the principal maximum for the average random realization of satellite positions (Fig. \ref{Pole_Radial_Densities} line (b)). Furthermore, we find that only $117$ of the $1000$ random realizations exhibited the degree of concentration of poles within $15^{\circ}$ of the principal maximum that was observed for the actual satellite distribution. Hence it would seem that a large percentage of satellite combinations are fitted by best-fit planes that all have strikingly similar orientations when compared with what one could expect from a random distribution of satellites. Again, this points toward a significant plane of satellites that includes a large fraction of the whole satellite sample.        

\begin{figure}[htbp]
\begin{center}
\includegraphics[width = 0.34\textwidth,angle=-90]{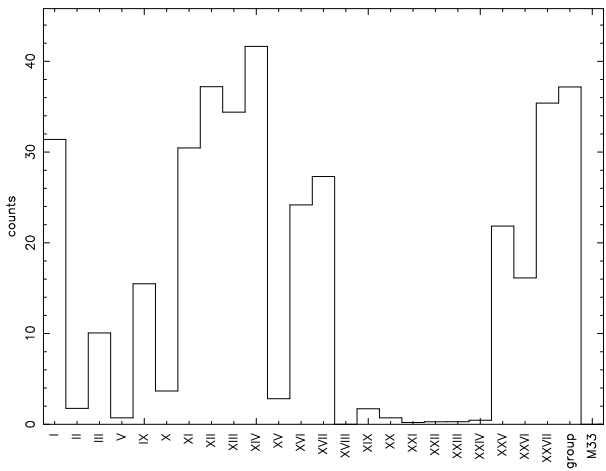}				  
\caption{Histogram showing the relative contribution of each satellite to the pole density within $3^{\circ}$ of the principal maximum at $l_{M31} = -78.7^{\circ}, b_{M31} = 38.4^{\circ}$. The histogram is generated from the same pole distribution as illustrated for $5$ satellites in Fig. \ref{Actual_Pole_Plots} and at the top of Fig. \ref{Random_Reals}.}
\label{sat_cont_prin_max}
\end{center}

\end{figure}

In order to obtain a better understanding of the satellites that this plane consists of, it is of particular interest to explore the number of times each satellite is included in a combination that is best fit by a plane with pole in close proximity to the principal maximum in the pole distribution for the entire sample. Once again, we use the pole distribution for all combinations of $5$ satellites, and we count the number of times each satellite contributes to a pole within $3^{\circ}$ of the principal maximum at $l_{M31} = -78.7^{\circ}, b_{M31} = 38.4^{\circ}$. The counts are divided by $100$ to account for the $100$ samples that are taken of each combination. The result can be seen in Fig. \ref{sat_cont_prin_max}. From this figure, it can be seen that the main contributors to the principal maximum in pole counts are those same satellites identified as forming a prominent plane in Fig \ref{sat_aitoff}, namely Andromedas I, XI, XII, XIII, XIV, XVI, XVII, XXV, XXVI, XXVII and the NGC147 group, along with Andromeda III and Andromeda IX. Hence the conclusion of our analysis thus far must be that there is indeed a significant plane in the satellite distribution of M31 and that it broadly consists of the aforesaid satellites. We therefore investigate the numerical significance of the best-fit plane to these satellites in \S \ref{Great_Plane}. As yet there is still more to be gleaned from a study of the pole density distribution however.        

\begin{figure*}[htbp]
\begin{center} 
$ \begin{array}{c}
\begin{overpic}[width = 0.50\textwidth]{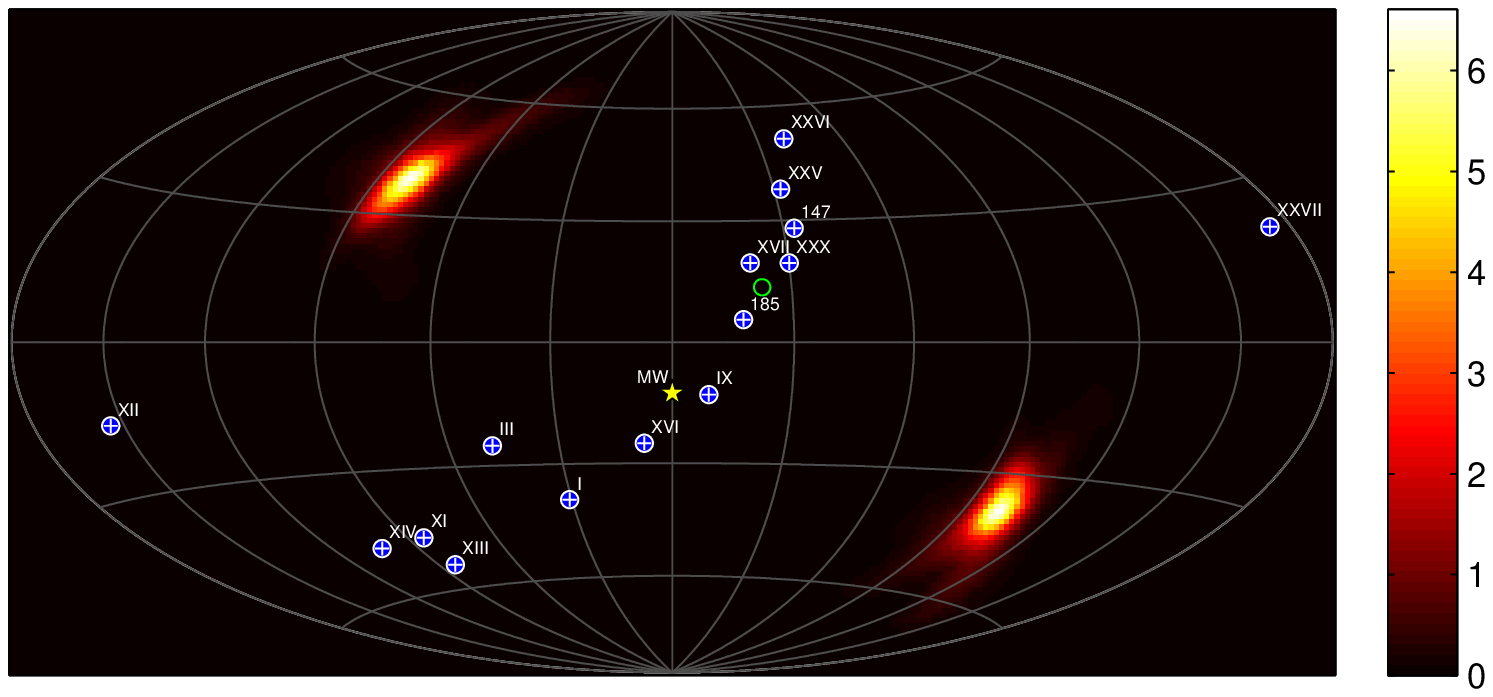}
\end{overpic}
\ \begin{overpic}[width = 0.50\textwidth]{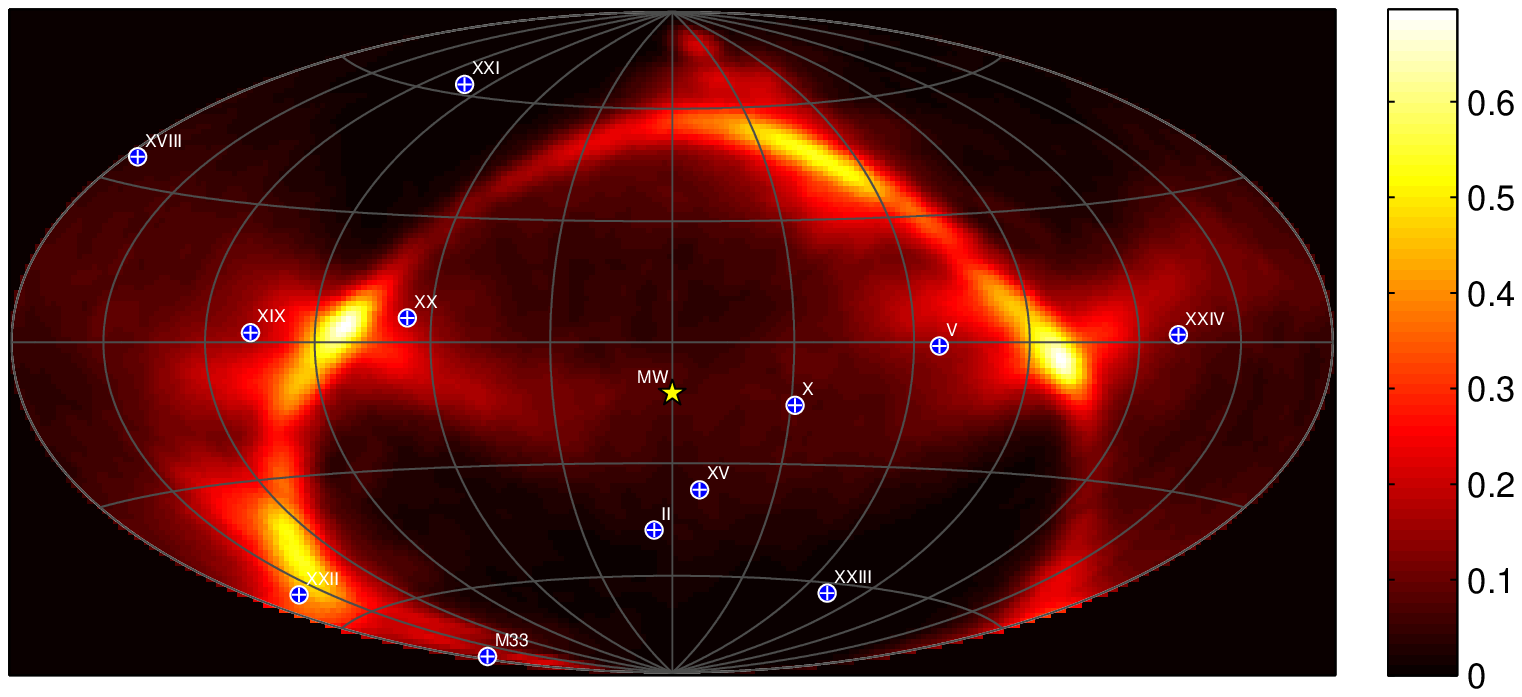}
\end{overpic}

\end{array}$
\end{center}
\caption{Pole density distributions generated from all combinations of 5 satellites possible from: Left) the satellites contributing significantly to the principal maximum at $l_{M31} = -78.7^{\circ}, b_{M31} = 38.4^{\circ}$ as per Fig. \ref{sat_cont_prin_max} and Right) the remaining $12$ satellites.}
\label{sat_cont_pole_plots}

\end{figure*}

From Fig. \ref{sat_cont_prin_max} we have been able to determine the principle contributing satellites to the principal maximum in the pole density distribution, but what of the remaining satellites? Do the positions of these satellites follow any particular trend? The best way to determine this is to construct pole density plots of the two halves of the complete sample, namely the major contributors to the principal maximum and the minor contributors. The resulting pole plots are presented in Fig. \ref{sat_cont_pole_plots}.   

The left-hand plot of Fig. \ref{sat_cont_pole_plots} shows the pole density distribution generated from the major contributing satellites to the principal maximum at $l_{M31} = -78.7^{\circ}, b_{M31} = 38.4^{\circ}$. This half-sample includes Andromedas I, III, IX, XI, XII, XIII, XIV, XVI, XVII, XXV, XXVI, XXVII and the NGC147 group. As expected, this plot reflects the existence of the aforementioned plane with all combination poles lying in the vicinity of the principal maximum. The right-hand plot, with poles generated from the remaining 12 satellites, namely Andromedas II, V, X, XV, XVIII, XIX, XX, XXI, XXII, XXIII, XXIV and M33, paints a very different picture however. There is a much greater spread in the distribution of poles, with the great circle induced by the survey area bias once again conspicuous. Also prominent in this plot are $2$ density maxima with their corresponding mirror images in the opposite hemisphere. The maximum lying midway between Andromedas XIX and XX lies very close to the pole of maximum detection bias at $l_{M31} = -90^{\circ}, b_{M31} = 0^{\circ}$ and so it is not unexpected, now that the prominent plane of satellites is effectively removed from the distribution. The elongated maximum passing through $l_{M31} \approx 45^{\circ},  b_{M31} \approx 45^{\circ}$ is more interesting however, and suggests the possibility of a second plane, roughly orthogonal to the major plane represented in the left-hand plot, though much less conspicuous. The planes represented by this maximum pass close to the error trails on the M31 sky of Andromedas II, III, XIX, XX, XXIII and XXIV. This maximum is faintly discernible in the pole distribution for combinations of $6$ satellites presented in Fig. \ref{Actual_Pole_Plots} but is no more pronounced than anywhere else along the high-density great circle in any of the other pole plots. On account of this, it would appear that this plane is likely no more significant than one would expect to find from a random satellite distribution subject to the same detection biases, such as those illustrated in Fig. \ref{Random_Reals}.   

\subsection{A Great Plane of Satellites}
\label{Great_Plane}

\begin{figure}[htbp]
\begin{center}
\includegraphics[width = 0.26\textwidth,angle=-90]{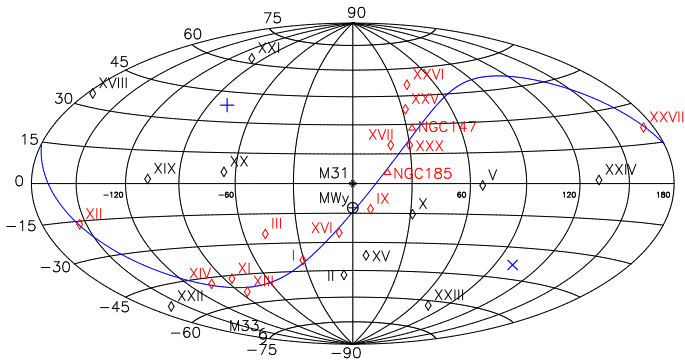}				  
\caption{A Great Plane of Satellites consisting of Andromedas I, III, IX, XI, XII, XIII, XIV, XVI, XVII, XXV, XXVI, XXVII, XXX, NGC147 and NGC185. The plane shown is that derived from the best-fit satellite positions. The pole is located at $(l_{M31}, b_{M31}) = (-78.4^{\circ}, 38.3^{\circ})$.}
\label{best_15_aitoff}
\end{center}

\end{figure}

Throughout the investigation undertaken thus far, all evidence has repeatedly pointed toward a conspicuously planar sub-set of satellites consisting of roughly half the total sample of satellites. Andromedas I, XI, XII, XIII, XIV, XVI, XVII, XXV, XXVI, XXVII and XXX as well as the dwarf ellipticals NGC147 and NGC185 all appeared to lie along a plane in Fig. \ref{sat_aitoff}. The reality of this co-planarity was verified in \S \ref{sat_subsets} and in particular Fig. \ref{sat_cont_prin_max}, which also suggested that Andromeda III and Andromeda IX should be considered as plane members. Hence it is of great interest to ascertain whether this `great plane' is in fact significant. To do this, it is necessary to determine how likely such a plane is to arise from a random satellite distribution subject to the same selection biases. The plane itself and the satellites of which it is constituted are illustrated in Fig. \ref{best_15_aitoff}. The plane shown is that calculated from the best-fit satellite positions and has equation of the form: $0.158x + 0.769y + 0.620z = 0$ with pole at $(l_{M31}, b_{M31}) = (-78.4^{\circ}, 38.3^{\circ})$. Note that for this section, we re-instate NGC147, NGC185 and Andromeda XXX as separate objects since we are again concerned with measurements of the significance of the planarity of the distribution. Our `great plane'  thus consists of $15$ satellites out of the entire sample of $27$. 

\begin{figure}[htbp]
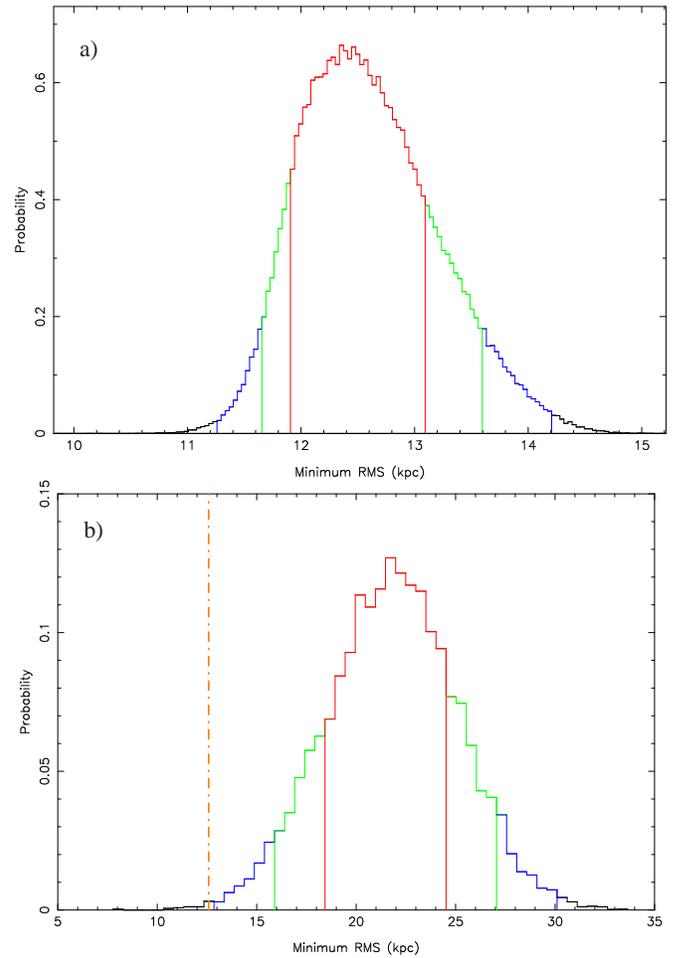

\begin{center} 
$ \begin{array}{c} 
\begin{overpic}[width = 0.35\textwidth,angle=-90]{Figures/err_samp_Best_15.ps}
\put(10,65){\small a)}
\end{overpic}
\\ \begin{overpic}[width = 0.35\textwidth,angle=-90]{Figures/Best_15_Sats_sig.ps}
\put(10,65){\small b)}
\end{overpic}
\end{array}$
\end{center}
\caption{Determining the significance of the observed `great plane' of satellites (see Fig. \ref{best_15_aitoff}). Figure (a) gives the distribution of possible values of the RMS obtainable from 200,000 realizations of possible positions of the $15$ plane members, given their respective distance probability distributions. Figure (b) plots the \emph{average} RMS of the best fit plane through the most planar combination of $15$ satellites for each of $10,000$ random realizations of $27$ satellites. These satellites are subject to the same selection biases as the real data. As for Fig. \ref{BFP_PPDs}, histogram (b) should only be compared with the average of histogram (a), which is plotted in (b) as a dashed line. It is thus clear that the planarity observed for our `great plane' of satellites is very unlikely to arise by chance. The $1 \sigma$ ($68.2 \%$), $90 \%$ and $99 \%$ credibility intervals are shown as red, green and blue lines respectively.}
\label{best_15_PPDs}

\end{figure}

Using the method of \S \ref{ss_Method_rr}, we again generate $10,000$ independent random realizations of $27$ satellites and seek the most planar combination of $15$ satellites from each. For each random realization, we sample $1000$ possible positions for each satellite as in previous sections and take the average value for the RMS of the best fit plane through the most planar combination. Since there are more than $17$ million ways that $15$ satellites can be drawn from $27$, and since we are not concerned with the orientation of each fitted plane as we have been in all previous sections, we depart from the plane fitting method of \S \ref{ss_Method_pf} for this section and instead proceed as follows. For each sample of satellite positions from each realization, $10,000$ randomized planes are generated and the $15$ closest satellites of the $27$ to the plane are stored in each case and the RMS recorded. The lowest RMS achieved is hence taken to be that for the most planar combination of 15 satellites in the sample. These minimum RMS values from each of the $1000$ samples of the particular random realization are then averaged to provide the best representation for the realization, given the distance uncertainties. Fig. \ref{best_15_PPDs} provides probability distributions in the RMS for the observed `great plane' $(a)$ together with those for the average RMS for the most planar combination from each random realization $(b)$. The average RMS for the observed plane is plotted in (b) for comparison. 

As can be seen from Fig. \ref{best_15_PPDs}, the RMS for the observed plane is very low compared to what one could reasonably expect from a chance alignment. Indeed, the average RMS of $12.58$ kpc for the observed plane is found to be equalled or exceeded in only $36$ out of the $10,000$ random realizations. The chances of obtaining such a planar group of 15 satellites from a sample of $27$ at random is thus estimated as $0.36 \%$. Hence we can conclude from this test that the observed plane is very unlikely to be a chance alignment, but rather the result of some underlying physical mechanism. Note that an independent but equivalent investigation is presented in ILC13 where such an alignment is found to occur in only $0.15 \%$ of instances. This is due to the larger central obstruction adopted in that analysis (19.6 vs. 7.9 sq. deg.) which rejects more satellites in close proximity to the plane pivot point (M31) where small plane distances are most likely.


\section{Discussion}
\label{s_Discussion}

Throughout the analysis conducted in \S \ref{s_results}, the presence of a prominent plane of satellites has been a consistent feature. This is not the first time that a significant plane of satellites has been identified from among the denizens of the M31 halo however. As early as 1995, \citet{Fusi95} had noted that the majority of the then known satellites lay conspicuously in a plane oriented edge-on with respect to the Milky Way. \citet{Koch06} identified a highly significant plane lying within $5^{\circ}$ to $7^{\circ}$ of being polar. Furthermore, they identify a subset of $9$ satellites from this plane lying within a thin disk with an RMS of $16$ kpc. \citet{Metz07} and later \citet{Metz09B} similarly identify a disk of satellites, this time not so markedly polar, with pole (in our coordinate system) at $(l_{M31}, b_{M31}) = (-70.2^{\circ}, 32.9^{\circ})$. They find this disk to have an RMS height of $39.2$ kpc. This disk is clearly the same structure that we identify here, being tilted by only $8.6^{\circ}$ with respect to our `great plane.' Our plane is found to have a much smaller RMS of just $12.34^{+0.75}_{-0.43}$ kpc however, despite including a comparable number of satellites. It is particularly noteworthy however, that their satellite sample is significantly different to that used here, with their disk including M32, NGC205, IC10, LGS3 and IC1613 - all of which lie outside the portion of the PAndAS survey region used in this study (see Fig. 10 (c) of CIL12). Indeed, it is clear from Fig. 4 of \citet{McConn06} that the galaxies M32, IC10, LGS3 and IC1613 all lie along the same great circle as our `great plane' in Fig. \ref{best_15_aitoff}, as do their entire error trails. Their conformity along with Andromeda I to a thin disk is noted in the said paper as one of $8$ possible `streams of satellites,' thus providing another early detection of the plane identified by this study. Both \citet{Majewski07} and \citet{Irwin08} also draw attention to the linear distribution of many of the plane-member satellites on the sky, a consequence of the edge-on orientation of the plane as indicated by the present study. The plane of \citet{Metz09B} does however include a significant number of satellites that, whilst included in our sample, we exclude due to their looser association with our plane. This then accounts for the much smaller RMS height observed in our study. 

Unlike previous studies of the M31 satellite system, we have a significant advantage in this study on account of the greatly improved sample of satellites available to us. Our sample is not only more numerous, but the positions are all determined via the same method applied to the same data as per CLI11 and CIL12. We are thus afforded unprecedented knowledge of the satellite detection biases, as well as the uncertainties in the object positions and have factored this knowledge into the analysis. An understanding of this bias is of particular importance when it comes to ascertaining the significance of any substructure identified, since a physically homogeneous satellite distribution will inevitably appear anisotropic after `folding in' the selection function and it is important that we do not attribute physical significance to this anisotropy.

Even after taking these effects into account however, there can be little doubt that the plane described in \S \ref{Great_Plane} is a real physical object. The component satellites extend well into the regions of low detection bias in Fig. \ref{survey_mask_effects} and the analysis of the last section makes it clear that such a thin disk of satellites has very little chance of arising within a random satellite distribution of the same size, even when subject to the same observation biases. Furthermore, it should be noted that the study of the plane's significance in \S \ref{Great_Plane} is likely to be conservative, given that if the satellites M32, IC10, LGS3, IC1613 and NGC205 were to be included in the analysis, the significance of our observed plane would likely grow still further. What is also particularly interesting is that subsequent research has shown $13$ of the $15$ objects to be co-rotating. This result is discussed in more detail in ILC13.   

What then could be the progenitor of this `great plane'? The polar orientation one might expect to arise had the satellites formed within the dark matter halo or had the dynamical friction proposed by \citet{Quinn1986} had sufficient time to take effect is not observed. Similarly, the findings of \citet{Metz09} seemingly preclude the possibility that the structure might be the result of the accretion of an external galactic association. Furthermore, there is apparently no marked distinction in the metallicities of the disk members compared with the non-disk members as one might expect from this scenario, though it is possible that some of the non-disk satellites may share the same origin as the disk members. There remains however the possibility that the satellites trace out the tidal debris of a galaxy merger. This is a particularly interesting possibility, especially since the plane, when projected onto the M31 tangent plane, is in close alignment with the Giant Stellar Stream. Indeed, \citet{Hammer10} show that the Giant Stellar Stream could feasibly be the product of a major merger event that began around $9$ Gyr ago, sustained by the returning stars from a tidal tail oriented similarly to our `great plane.' 

If a link is to be established, the observed asymmetry of the system must also be compatible with the tidal scenario however. It is of particular interest that, of the $13$ co-rotating satellites in the plane, all but one lie on the near side of the M31 tangent plane. Indeed, if we removed all of the plane member-satellites from the system, the remaining satellite distribution would no longer be significantly asymmetric. With almost all of the satellites currently on the near side of M31, it would seem on first consideration that the progenitor event could not have occurred substantially more than a typical orbital time ago or else the satellites would have had sufficient time to disperse. This suggests the event responsible must have occurred within the last $5$ Gyr. Other studies however have supported the proposition that a group of tidal dwarf galaxies could survive for extended periods whilst retaining the asymmetry inherent from the time of their formation. For instance, \citet{Duc11} find three moderately old tidal dwarf candidates with this asymmetry preserved. It must also be remembered that the precise orbits of the satellites are undetermined and so it is not clear how close they have come to M31 in the past. As discussed by \citet{Paw11}, there is also evidence that tidal material can survive in a 'bridge' between the interacting galaxies for an extended period. Nevertheless, it must be cautioned that there is no established precedent for tidal dwarfs with the longevity implied by the stellar populations of the M31 satellites, and the fact that they continue to adhere to such a thin structure is even more perplexing under such a scenario.


There is also another striking characteristic of the observed plane. As one will note from examination of Fig. \ref{best_15_aitoff} (and indeed the left-hand column of plots in Fig. \ref{Actual_Pole_Plots}), it is oriented perfectly edge-on with respect to the Milky Way. Whilst there is a noted bias toward detection of satellites positioned along planes oriented in this way, it must be remembered that this bias arises primarily due to the propensity for detecting satellites close to the line of sight passing through M31. Many of the satellites observed to lie on our plane are located a good distance from this line of sight however and well into the low-bias portions of the M31 sky. In any case, the random realizations of \S \ref{Great_Plane} suffer from the same biases and yet show unequivocally that the observed plane is very unlikely to arise by chance. Hence if we are to accept these results, we must also accept the plane's orientation. 

Further to this strikingly edge-on orientation, it is also noteworthy that the plane is approximately perpendicular to the Milky Way disk. This fact can be easily seen if the constituent satellites are traced out in Galactic coordinates (i.e. all lie on approximately the same Galactic longitude). This of course raises the question - how does the orientation of the Milky Way's polar plane of satellites compare with this plane? Noting that the average pole of the `Vast Polar Structure' described by  \citet{Paw12B} points roughly in the direction of M31, the two planes are approximately orthogonal. These precise alignments are discussed in more detail in ILC13, but suffice to say here that this alignment is particularly interesting and suggests that the Milky Way and M31 halos should not necessarily be viewed as fully isolated structures. It is entirely conceivable that our current ignorance as to the coupling between such structures may be to blame for our inability to pin down the precise mechanism by which such planes arise.

Finally, in consequence of these findings, most particularly with respect to the highly significant, very thin disk of satellites that has been identified, it is clear that if $\Lambda CDM$ is to remain the standard model of cosmology, the occurrence of such structures has to be explicable within it. The possible deeper implications of the satellite anisotropies are discussed in \citet{Kroupa10}, \citet{Angus11}, \citet{Fouquet12} and \citet{Kroupa12} wherein other alternative cosmologies are also highlighted.



\section{Conclusions}
\label{s_Conclusions} 
It is clear that whilst the satellites of M31 when taken as a whole are no more planar than one can expect from a random distribution, a subset consisting of roughly half the sample \emph{is} remarkably planar. The presence of this thin disk of satellites has been conspicuous throughout the analysis contained in this paper. The degree of asymmetry determined from the satellite distribution is also found to be relatively high. Of particular note, the orientation of the asymmetry is very significant, being aligned very strongly in the direction of the Milky Way. When this fact is combined with the apparent orthogonality observed between the Milky Way and M31 satellite distributions and the Milky Way disk, it appears that the two halos may in fact be coupled. Regardless, the great plane of satellites identified in this study, and its clear degree of significance, is not directly expected from $\Lambda CDM$ cosmology. This finding provides strong evidence that thin disks of satellites do indeed exist in galaxy halos, and whether or not the standard model can account for such structures remains to be seen.


\begin{acknowledgments}
A. R. C. would like to thank Macquarie University for their financial support through the Macquarie University Research Excellence Scholarship (MQRES) and both the University of Sydney and Universit{\'e} de Strasbourg for the use of computational and other facilities.
G. F. L. thanks the Australian Research Council for support through his Future Fellowship (FT100100268) and Discovery Project (DP110100678). 
R. A. I. and D. V. G. gratefully acknowledge support from the Agence Nationale de la Recherche though the grant POMMME (ANR 09-BLAN-0228). Based on observations obtained with MegaPrime/MegaCam, a joint project of CFHT and CEA/DAPNIA, at the Canada-France-Hawaii Telescope (CFHT) which is operated by the National Research Council (NRC) of Canada, the Institut National des Science de l'Univers of the Centre National de la Recherche Scientifique (CNRS) of France, and the University of Hawaii.
\end{acknowledgments}



\end{document}